\documentclass[fleqn,10pt]{wlscirep}
\usepackage[utf8]{inputenc}
\usepackage[T1]{fontenc}
\usepackage{amsmath}
\usepackage{graphicx}
\usepackage[colorlinks=true, allcolors=blue]{hyperref}
\usepackage{booktabs}
\usepackage{subcaption}
\usepackage{multicol}
\title{Load Balancing For High Performance Computing Using Quantum Annealing} 

\author[1]{Omer Rathore}
\author[1]{Alastair Basden}
\author[1,2]{Nicholas Chancellor}
\author[1,3]{Halim Kusumaatmaja}
\affil[1]{Department of Physics, Durham University, Durham, DH1 3LB, UK}
\affil[2]{School of Computing, Newcastle University, Newcastle upon Tyne, NE4 5TG, UK}
\affil[3]{Institute for Multiscale Thermofluids, School of Engineering, The University of Edinburgh, Edinburgh, EH9 3FB, UK}

\affil[*]{omer.rathore@durham.ac.uk, a.g.basden@durham.ac.uk, nick.chancellor@newcastle.ac.uk, halim.kusumaatmaja@ed.ac.uk}

\begin{abstract}
With the advent of exascale computing, effective load balancing in massively parallel software applications is critically important for leveraging the full potential of high performance computing systems. Load balancing is the distribution of computational work between available processors. Here, we investigate the application of quantum annealing to load balance two paradigmatic algorithms in high performance computing. Namely, adaptive mesh refinement and smoothed particle hydrodynamics are chosen as representative grid and off-grid target applications. While the methodology for obtaining real simulation data to partition is application specific, the proposed balancing protocol itself remains completely general. In a grid based context, quantum annealing is found to outperform classical methods such as the round robin protocol but lacks a decisive advantage over more advanced methods such as steepest descent or simulated annealing despite remaining competitive. The primary obstacle to scalability is found to be limited coupling on current quantum annealing hardware. However, for the more complex particle formulation, approached as a multi-objective optimization, quantum annealing solutions are demonstrably Pareto dominant to state of the art classical methods across both objectives. This signals a noteworthy advancement in solution quality which can have a large impact on effective CPU usage.
\end{abstract}

\begin{document}
\flushbottom
\maketitle
%
%
\thispagestyle{empty}

\section*{Introduction}
During the initial development of scientific computation, simulations greatly benefited from improvements to clock speeds of individual processors \cite{danowitz2012cpu}. However, since the early 2000s \cite{sutter2005free} hardware limitations have resulted in dwindling improvements to sequential programming applications. The result has been a paradigm shift to concurrency in programming software that aims to exploit the multi-core architectures which form the bedrock of modern day high-performance computing (HPC). 
However, the effectiveness of these applications is heavily dependant on the equitable distribution of computational workload across available resources, a concept known as load balancing. 


Load balancing encompasses not just fair distribution of the computational tasks across processors, but also doing so in such a way that minimises the need to communicate data between processors. Motivation for the former is evident considering how all processors need to wait for the slowest one to finish before proceeding to the next step, thus potentially resulting in idling and resource wastage at a much larger scale if even a single processor lags behind. In addition to this, communication bandwidth between processors is usually not as efficient as within the processor itself \cite{tan2009influence,rabenseifner2003hybrid} and ideally should be minimised.
The importance of load balancing clearly extends beyond any singular discipline. 
However for the purposes of this paper emphasis is placed on applications in the realm of computational fluid dynamics. 
In particular, this paper explores the viability of quantum annealing as a solution strategy, as demonstrated using representative grid and off-grid applications. 

In the context of grid based applications, it is usually not the governing equations themselves that define the load balancing problem but the chosen method of domain decomposition (DD). DD is the art of splitting a computational domain into smaller sub-domains. This allows solving each smaller problem individually before subsequently recombining into the global solution. Historically, since its original proposal \cite{schwarz1869ueber}, DD in the realm of fluid simulations has been used to accommodate the inherent challenges of complex geometries \cite{hessenius1982zonal,henshaw1987multigrid} or higher dimensionality \cite{glowinski1983domain}. Extensive research in the last 40 years has led to rapid development, facilitating application to multiphysics problems \cite{keyes2013multiphysics}, moving meshes \cite{houzeaux2017domain} and a wider pool of applications \cite{tang2021review}. More importantly in the context of modern HPC, DD is an indispensable tool for transforming the global simulation domain into sub-components that can be assigned to individual processors. 

A popular choice of DD for large scale simulations in the realms of fluids \cite{wang2010adaptive}, stress analysis \cite{bennett1985structural} and biological flows \cite{botti2010adaptive} among others is adaptive mesh refinement (AMR). AMR operates on the principle that different areas of the computational domain are best solved with different levels of spatial resolution. This can prevent wastage of computational resources in regions with little activity while maintaining precision in zones of interest. The method is particularly useful for applications with large variation of spatial scales or when high resolution data is required in specific regions. 
This paper considers block structured implementations as a representative class and does not make further distinctions with alternative AMR varieties, although an interested reader is referred to the relevant literature \cite{zumbusch2012parallel,dubey2014survey}. 

For off-grid applications, the representative method here is smoothed particle hydrodynamics (SPH), a mesh free simulation method for continuum mechanics. Since its original conception \cite{gingold1977smoothed}, its application has spread to encompass a wide variety of fields including but not limited to solid mechanics, engineering, astrophysics and the food industry \cite{monaghan2012smoothed,shadloo2016smoothed}. 
Fundamentally, SPH is an interpolation technique that utilizes a collection of unordered points, or particles, to ascertain the value of a function at a specific point. This process is facilitated through the use of a kernel function, which effectively integrates the influence of adjacent particles to compute the value of a function at the desired point.

When applied in the context of computational fluids this forms a Lagrangian particle method for solving hydrodynamic equations, thus forming an equivalent counterpart to the Eulerian equations solved by grid based methods. Moreover, when compared with the latter, SPH is inherently more adept at handling free surface flows and complex boundaries and can accommodate large deformations without being hindered by grid distortion. However, these advantages are usually accompanied with caveats when it comes to stability and convergence, as well as limited accuracy \cite{lind2020review}. 

This paper delves into the potential of quantum computing to address the challenges of load balancing, with a particular focus on the quantum annealing (QA) approach \cite{kadowaki1998quantum} for each of the two cases described above.
QA is particularly suited for finding the ground state of an Ising problem, which is essentially analogous to finding the optimal solution to many binary combinatorial optimisation problems of interest. This includes the travelling salesman and knapsack problem, as well as more niche applications in computational chemistry or graph theory \cite{lucas2014ising,camino2023quantum,ushijima2017graph}. 
In light of its range of applicability, the field has attracted considerable interest in recent years, a trend that has only been further fueled by the increasing accessibility of quantum annealers—programmable quantum computers designed for quantum annealing. While the largest gate based quantum computer (IBM) only has 433 quantum bits (i.e. qubits) \cite{collins2022ibm}, the largest annealer (D-Wave) has over 12000 qubits \cite{pressrelease2023}. 

There is an expectation in the community that QA might outperform classical algorithms for some applications. This has led to many empirical studies into the efficacy of QA for a wide range of problems \cite{katzgraber2014glassy,katzgraber2015seeking,boixo2016computational,denchev2016computational,yarkoni2022quantum}, often with mixed results that lack a decisive consensus. As of yet, there may be more promise in using QA for approximate optimisation based on a recent demonstration of a scaling advantage over the best classical algorithm, even though this was for a rather artificial optimisation problem \cite{bauza2024scaling}. 

As such it seems the actual choice of problem to solve strongly influences any inferences made about the potential (or lack thereof) of quantum annealing. While there are many works in the existing literature that apply QA to a particular problem, the body of literature tackling the more fundamental question of whether QA even should be applied or not, is very sparse. 
A recent attempt \cite{chancellor2020toward} to create a rigorous methodology for evaluating the viability of a potential use case for quantum computing aims to shed some light on the latter. However it should be noted that the discussion remains very open ended, despite being of paramount importance particularly in the context of near term quantum hardware. 

So why does this paper consider load balancing as the target application for QA? 
Firstly, it should be evident that load balancing is a topic of relevance for almost all computational scientists regardless of discipline. This is particularly true as the exascale era of classical computing approaches and programmers strive towards utilising more and more compute power. 
It is also still very much an area of active research \cite{mena2022transparent,zhu2023novel,mohammed2020two,miller2021dynamic}, highlighting that the room for potential improvements to be made persists. 
Furthermore, as will be seen shortly, the problem lends itself to a very natural conversion into an Ising formulation suitable for quantum annealers. Although it is not always inherently apparent \emph{a priori} how complex the solution energy landscape will be, it is at the very least very large and scales drastically with problem size. 

Perhaps the most compelling rationale for selecting load balancing as an application for QA emerges when considering the broader HPC landscape. 
Classical computing, benefiting from several decades of development, significantly outpaces the emergent field of quantum computing in terms of maturity and scale when it comes to hardware. Despite the immense potential of quantum technology, it's unlikely to match the sheer scope of classical systems in the near future. Additionally, there's a growing consensus that classical computing will always retain some relevance, never being completely supplanted by its quantum counterpart.  
In this context, envisioning quantum computers as complementary accelerators, rather than standalone solutions, is a strategic and viable approach. Particularly in load balancing, where recalculations aren't constant but occur at regular or semi-regular intervals and hence quantum annealers can effectively augment the process.
The proposal here is to integrate quantum annealers with classical HPC systems. The primary computational tasks, whether in fluid dynamics, solid-state mechanics, biological flows or some other algorithm, continue to run on classical HPC infrastructure. 
Meanwhile, the load balancing component is offloaded to an attached quantum processor operating in parallel. 
This concept mirrors the synergy between GPUs and CPUs \cite{liu2012accelerating}, and aligns well with the growing narrative of heterogeneous HPC architectures. 
Such an approach not only capitalizes on the strengths of quantum annealers but is also in tune with the trend \cite{chen2009high,tiwari2015understanding,Callison2022Hybrid} towards diversifying and optimizing computational resources in HPC environments. 

The structure of the remainder of this paper is outlined as follows: firstly, an introductory review of quantum annealing is presented, alongside pertinent details from the classical methodologies under consideration. 
This encompasses an in-depth discussion on the acquisition of real-world classical data to be subsequently load balanced using QA. 
After which, the outcomes derived from both grid-based (AMR) and off-grid (SPH) simulations run on actual quantum annealing hardware are explored.
\section*{Methods}

\subsection*{Quantum Annealing}
Of the myriad possibilities \cite{de2021materials} when it comes to implementing a quantum computer, currently the two leading paradigms are gate based quantum computing (QC) \cite{kwon2021gate,nielsen2010quantum} and adiabatic quantum computing (AQC) \cite{albash2018adiabatic,van2001powerful}. 
In principle the two approaches have been shown \cite{aharonov2008adiabatic} to be equivalent if allowing a polynomial overhead. 
Gate based QC can be thought of as the quantum analogue to classical computing where instead of logical gates acting on bits, the computation is performed by applying unitary gates to qubits. 
Alternatively, AQC relies on adiabatic time evolution of a quantum state from an initial, easy to prepare state to a final observed value as modelled by the Schr\"odinger equation.

Originally proposed as a way to tackle satisfiability problems \cite{farhi2000quantum}, AQC has since received significant attention since due to the ease with which many combinatorial optimisation problems can be represented in Hamiltonian form \cite{lucas2014ising}.
The protocol operates under the premise of the adiabatic theorem \cite{kato1950adiabatic}, which states that a quantum system in its ground state remains in its ground state when acted upon by a perturbation if the changes to the Hamiltonian are slow enough and there is an energy gap between the ground and excited states. Therefore if an initial Hamiltonian is appropriately evolved into the problem Hamiltonian the final ground state should encode the solution to the desired problem. This is usually done by interpolating between a Hamiltonian with an easy to directly prepare ground state ($H_A$) and the problem Hamiltonian ($H_B$) to give the instantaneous value as described in Equation \ref{e1}. 
\begin{equation}\label{e1}
H(t)=A(t)H_A + B(t)H_B    
\end{equation}
The temporal prefactors ensure a smooth transition between the two Hamiltonians. Initially $H_A$ dominates ($A(0)=1, B(0)=0$), prior to ceding to $H_B$ over time ($A(T)=0,B(T)=1$). The rate of this change is crucial in preventing disruptive, diabatic changes to the system and the acceptable limits are dependant on the minimum energy gap of the problem \cite{jansen2007bounds,elgart2012note}. 

A common choice for the initial Hamiltonian is a transverse field in the x direction,
\begin{equation}\label{e2}
    H_A=\sum_{i\in V} \sigma_i^x,
\end{equation}
with $\sigma^x_i$ being the x-Pauli matrix acting on qubit "i". 
This is eventually replaced by the problem/final Hamiltonian as described by,
\begin{equation}
    H_B= \sum_{i\in V} h_i\sigma_i^z +\sum_{(i,j) \in E} J_{ij} \sigma_i^z \sigma_j^z ,
\end{equation}
where $G(V,E)$ is the graph consisting of qubit nodes $V$ and connective edges between neighbouring qubits $E$. This Hamiltonian is characterised by the local field at the i-th qubit, represented by $h_i$, as well as the interaction couplings between connected qubits denoted by $J_{ij}$.
The instantaneous Hamiltonian during a computation, $H(t)$, of Equation \ref{e1} is thus the well known transverse-field Ising Hamiltonian \cite{stinchcombe1973ising}. As the contribution of $H_A$ fades over time, the quantum dynamics fade out and  eventually result in a purely classical system where the qubits are measured in order to obtain the solution. The classical Ising model can be readily obtained by replacing the $\sigma^z$ Pauli operators with classical spin variables : 
\begin{equation}
    H_{Ising}=\sum_i h_is_i + \sum_{ij}J_{ij}s_is_j .
\end{equation}

Moreover by operating under the umbrella of the adiabatic theorem, the protocol is guaranteed to obtain the ground state solution, at least in theory. However, in practice the requirements of true adiabacity and changing the Hamiltonian slowly enough can be excessively stringent. This is clearly evident in current architecture where  thermal coupling with the environment is strong enough such that the timescales are of the order $\sim10-100$ns \cite{king2022coherent}. Furthermore the requirements on how quickly the Hamiltonian can be evolved (i.e. the annealing schedule) is dependant on the gap between the ground state and first excited solution, which is not usually known \emph{a priori}. 

QA is thus a relaxation of these conditions that follows the same principles as AQC, but forms a distinctly separate algorithm, with the annealing schedule determined heuristically and strictly adiabatic conditions not guaranteed.
The same time dependant Hamiltonian of Equation \ref{e1} is used to evolve the quantum state, however the system can no longer accurately be modeled by the Schr\"odinger equation, but would require master-equation based descriptions.
As a result, QA forms a heuristic quantum algorithm that should be viewed as a statistical sampler rather than a deterministic solver, with the intended goal of maintaining a relatively high probability of remaining in the ground state or to at least be sufficiently close. The theory of quantum annealing is much less developed than AQC, partially due to the more complex setting. However, some progress has been made, for example in the topic of diabatic computing which considers rapid quenches far from the adiabatic limit \cite{Crosson2021Diabatic}. In particular, in this regime a mechanism related to energy conservation provides an important guarantee on average optimality \cite{Callison2021Energetic}. 

There are actually many similarities between QA and a well known \cite{bertsimas1993simulated} classical counterpart known as simulated annealing (SA). The latter aims to mimic the physical process of a solid being slowly cooled so that the eventual ``frozen'' state contains the lowest energy solution to a desired cost function. In essence the approach relies on thermal fluctuations for exploration of the solution landscape while avoiding local minima. Meanwhile QA takes a very thematically similar approach but replaces thermal fluctuations with quantum ones. $H_A$ introduces disorder into the system with respect to the ground states of $H_B$ and in doing so provides qubits with energetic variations during the anneal.

\begin{figure}[!h]
    \centering
    \includegraphics[width=0.5\linewidth]{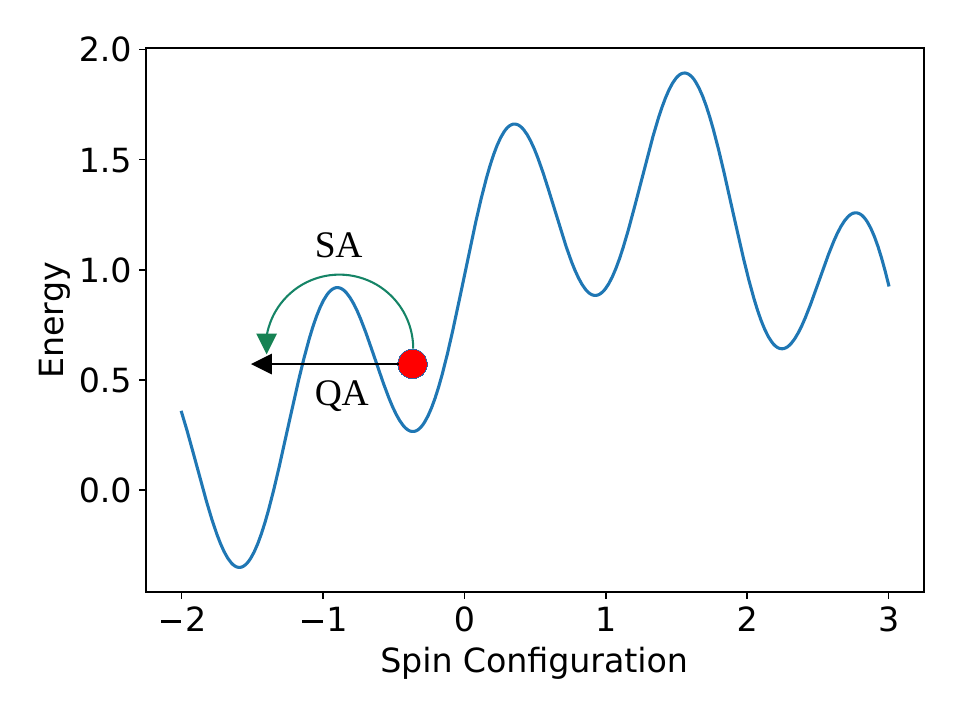}
    \caption{When in a local minima, QA can use quantum tunneling to directly escape. SA instead relies on thermal fluctuations to escape over the energy barrier.}
    \label{fig:tunneling}
\end{figure}

The expectation that Quantum Annealing (QA) might surpass classical algorithms is tied to the potential impact of quantum mechanical phenomena such as superposition, tunneling, and entanglement \cite{albash2018adiabatic,yarkoni2022quantum,yaacoby2022comparison,starchl2022unraveling,Chancellor2021Range}. Fundamentally, quantum superposition and tunneling can enable transitions between states, even those separated by high energy barriers \cite{razavy2013quantum}. This suggests that a search algorithm utilizing QA could overcome local minima more easily by tunneling through these energy barriers as illustrated in Figure \ref{fig:tunneling}.

QA is thus grounded in a model that more closely reflects the behavior of real quantum systems when compared to AQC. However, the potential for non-adiabatic effects complicates the distinction between its computational complexity and that of conventional computing. QA often involves the non-adiabatic evolution of mixed quantum states within an open system, and this approach is generally considered heuristic. Thus, it lacks the provable computational advantages that AQC provides \cite{grant2020adiabatic}.

The general approach to employing a quantum annealer shares many similarities across different architectures. This research, however, utilises D-Wave systems, which consist of superconducting qubits. Although there are several promising alternatives \cite{scholl2021quantum,lechner2015quantum,ebadi2021quantum}, these technologies are comparatively nascent in their development. Meanwhile D-Wave is the largest and most commonly used commercial QA platform currently and has been used extensively for both industry and research purposes. 
For a detailed, step-by-step guide on using a quantum annealer, interested readers are encouraged to consult the numerous introductory articles available on this topic \cite{yarkoni2022quantum,grant2020adiabatic,de2011introduction}. In summary, the primary steps involved in utilizing an annealer include problem formulation, minor embedding, and sampling (i.e. the anneal itself). Problem formulation and how the classical simulations were run is described next, with the influence of embedding and sampling discussed as part of the results. 
\subsection*{Adaptive Mesh Refinement}
In order to formulate load balancing for AMR as an Ising problem suitable for annealers, data was gathered using CompReal \cite{rathore2023flame}, a fully compressible, finite difference flow solver for the Navier-Stokes equations. The flow solver interfaces with a widely used general software framework for AMR applications called BoxLib \cite{bell2012boxlib,dubey2014survey}. 
BoxLib was designed as a platform upon which to build massively parallel software and has demonstrated good scalability on up to 100,000 cores \cite{zhang2016boxlib}. 
Research codes based on BoxLib are numerous and varied, but include the realms of astrophysics \cite{almgren2010castro,nonaka2010maestro}, computational cosmology \cite{almgren2010castro}, subsurface flow \cite{pau2009parallel} and combustion\cite{day2000numerical} among others.  

\begin{figure}[h!]
    \centering
    \includegraphics[width=0.5\linewidth]{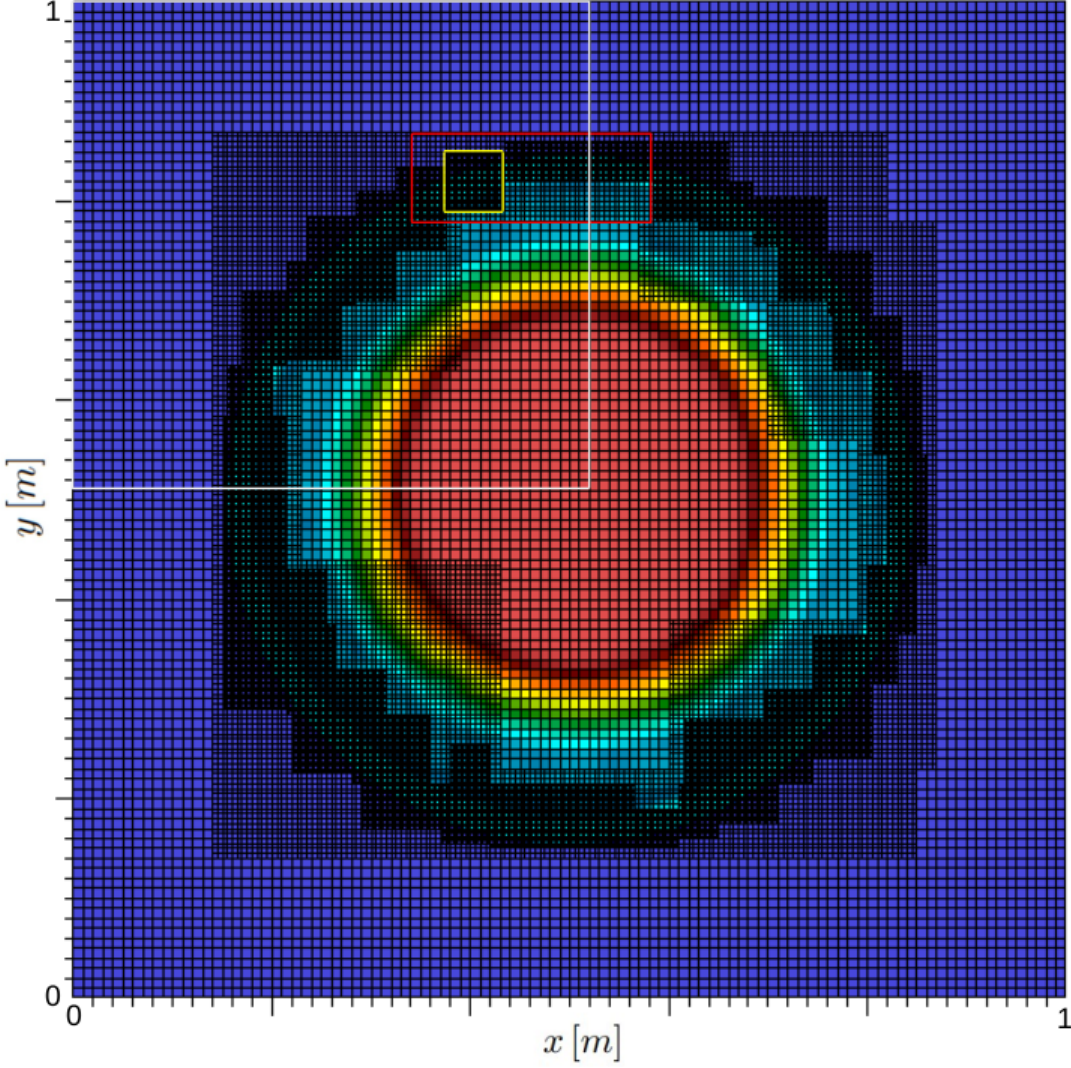}
    \caption{A shockwave expanding outwards from a region initially at high pressure (red) to the surrounding lower pressure (blue) environment simulated using CompReal. The nested grid hierarchy is illustrated using individual grids from the finest, intermediate and coarsest levels outlined in yellow, red and white respectively.}
    \label{fig:amr_gen}
\end{figure}

The particular test case simulated involves a developing spherical blast wave as illustrated in Figure \ref{fig:amr_gen}. The sharpest gradients are in a very thin zone encompassing the shock edge which is moving outwards. As such the mesh is regularly updated to track its position.  
Data is defined on a nested hierarchy of logically rectangular collection of cells called grids (or patches). This is divided into levels where each level refers to the union of all grids that share the same mesh spacing. Aside from the coarsest level, finer levels are disjoint and do not cover the entire simulation domain, thus facilitating allocation of resources to desired regions. Each grid is made up of a number of cells and this is not necessarily the same across all grids. The cells have high intra-connectivity within the same grid and lower inter-connectivity to neighbouring grids. Therefore, from a work distribution perspective, it is desirable to allocate whole grids to individual processors while trying to maintain roughly the same total cell count when possible.


To quantify the computational burden associated with each nested grid, the total number of cells within that patch are counted. There are also alternative measures of "burden", for example actually recording the time it takes each patch to complete its allocated tasks might be more representative for multi-physics simulations where different cells solve different equations. 
This would affect the value of the weights but not the Ising formulation itself so the method remains completely general.

The classical simulation thus provides a series of numbers representing the notional cost of each grid, which needs to be equitably distributed across a given number of processors. This process will need to be repeated throughout the classical simulation at certain intervals. It is undesirable to split these grids due to their high intra-connectivity and so the problem essentially reduces to one of number partitioning.
Given a set of N numbers, where N is the number of patches, $S=\{n_1,...,n_N\}$, the task is to divide this set into two disjoint subsets such that the sum in both elements is the same, or at least minimises the mismatch. This is framed as the following Ising model \cite{lucas2014ising},
\begin{equation}
H = A \left( \sum^N_{i=1}n_is_i \right)^2 ,  
\end{equation}
where $n_i$ are the numbers in the set, $s_i$ is the Ising spin variable and $A$ is a general scaling constant which is set to unity henceforth. Despite the absence of linear biases in the model, the coupling terms will lead to the formation of a fully connected graph when mapped to the quantum processor. It's worthwhile noting that number partitioning phrased as a decision problem regarding if the two subsets are equal or not is classified as NP-complete \cite{karp2010reducibility}. 
However, since this model is based on real data from simulations, achieving a perfectly balanced split is essentially almost impossible. Therefore, the objective shifts to minimizing the discrepancy between subsets, a task that is NP-hard. This process can be applied recursively to achieve divisions for more than two processors. Currently BoxLib uses a simple round robin strategy for distributing grids between subsets.
\subsection*{Smoothed Particle Hydrodynamics}
This paper is not concerned with the technical details behind discretising a set of governing equations using the SPH operator, as this is application specific and the interested reader can refer to the relevant literature \cite{lind2020review,monaghan2012smoothed}. 
Instead, the key takeaway here is the idea that this family of methods requires storage of particle data. Moreover these particles are irregularly positioned in space and move at each time step. Although this is a key strength of SPH as it allows particle aggregation in certain regions of interest and a sparse distribution in other areas, leveraging this benefit in practice is only viable with efficient memory and work allocation. 

Compact support of the kernel ensures that particles are only influenced by neighbours that fall within the smoothing length. So it would be logical to split the domain into sections when assigning work to processors in order to limit inter-processor communication. 
However as the particles are disordered it is important to remember some regions will be more densely populated than others, thus there is  also the second avenue of intra-processor communication to consider when creating a work distribution.   

Classical data to be partitioned, by using QA in this work, was obtained via SWIFT \cite{schaller2023swift}, a popular astrophysics code that has demonstrated good scalability on $100,000$ cores. SWIFT relies on task based parallelism \cite{schaller2016swift} in order to optimise shared-memory performance within each individual node as its backbone. This can then be readily generalised to create a work allocation across multiple, distributed memory nodes using graph partitioning algorithms.
In summary, the domain is initially divided into a set of \emph{cells}, with each cell containing a collection of particles such that if two particles interact they are either in the same cell or at most neighbouring cells. 

In terms of load balancing, this domain decomposition approach can be efficiently modeled as a graph, where each node symbolizes an individual cell containing some number of particles. The edges of this graph represent shared tasks reliant on particles located in separate cells that require communication between these cells. Moreover, tasks involving cells from different partitions must be processed by both respective processors, whereas tasks within the same partition are evaluated exclusively by the corresponding processor.
An optimal load balancing strategy aims to minimize both the sum of node weights for each subset and the sum of edge weights bridging these subsets. The former goal is to reduce the waiting time caused by the slowest processor at each step, while the latter focuses on decreasing the necessary bandwidth for inter-processor communication. This dual objective ensures both efficient processing and minimal communication overhead.

\begin{figure}
    \centering
    \includegraphics[width=0.5\linewidth]{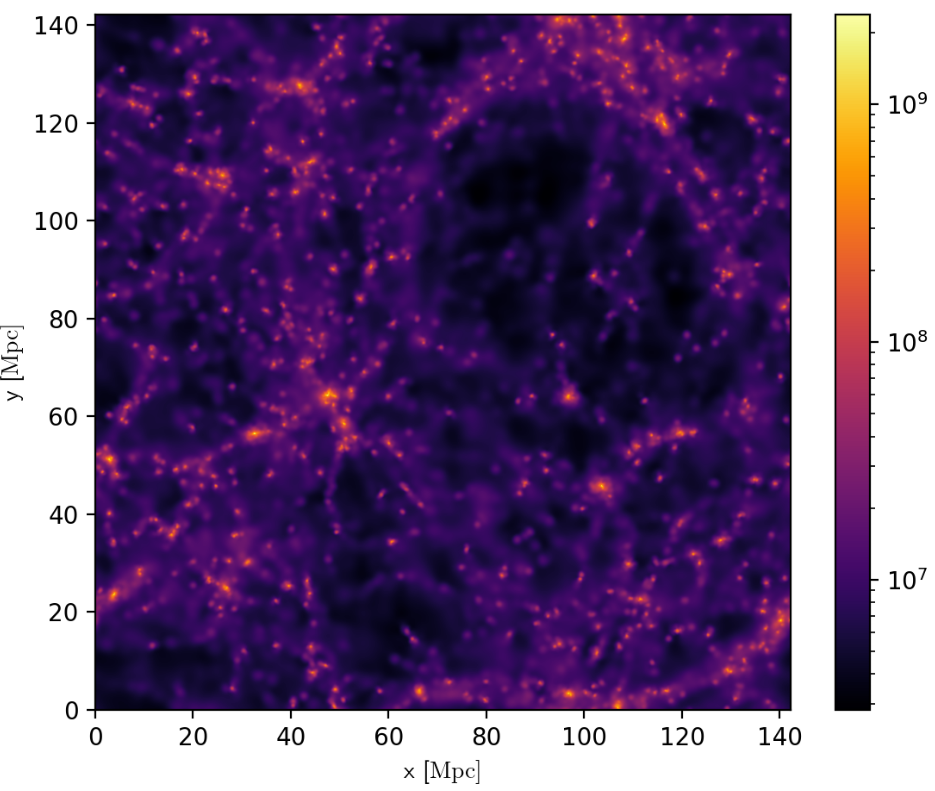}
    \caption{A snapshot at time $13.7$Gyr of the projected mass for a small cosmological volume with dark matter simulated using SWIFT. Note that due to local gravity wells some regions will have a higher density of particles than others.}
    \label{fig:swift_gen}
\end{figure}

Values for the weights were obtained by simulating a small cosmological volume with dark matter as found in the SWIFT example suite and illustrated in Figure \ref{fig:swift_gen}. The simulation involves $64^3$ particles and the weights were extracted by timing the intra/inter cell tasks respectively for node/edge weights. 
The domain decomposition parameters were modified by design to reduce the number of cells in order to accommodate the problem onto current annealers. Since the cells are cubes, in three dimensions this boils down to a minimum node/cell count of $27$ since each cell has 6 face adjacent, 12 edge adjacent and 8 corner adjacent neighbours. With all 26 neighbours residing within the 1 cell range defined earlier for shared tasks, this results in a fully connected graph. 

The Ising model \cite{lucas2014ising} for this load blancing problem places a spin variable ($-1/+1$) on each node in order to determine whether that node will eventually belong in the "$+$" or "$-$" subset. The energy functional consists of two components,
\begin{equation}\label{e99}
    H=\gamma H_1 + H_2,
\end{equation}
where a Lagrange parameter ($\gamma$) has been introduced to allow changing $\gamma$ in order to explore any conflicting effects. 
The model includes a penalty term when the node weight in set $+$ is not equal to that in set $-$,
\begin{equation}
    H_1=\left( \sum_{n=1^N} w_is_i \right)^2,
\end{equation}
where the sum is across all nodes, each with weight $w_i$. 
Additionally there is a penalty term for each time an edge that connects nodes in different subsets is cut, 
\begin{equation}
    H_2=\sum_{(uv)\in E}e_i\frac{1-s_us_v}{2},
\end{equation}
where the sum is across all edge connected nodes with $e_i$ referring to the weight of the edge. This is a NP-hard problem \cite{karp2010reducibility} of significant value in ensuring many particle based codes retain a meaningful advantage on future HPC systems. Currently SWIFT relies on the graph partitioning software METIS \cite{karypis1997metis}. 
\section*{Results}
\subsection*{Grid Based Application}
In the following discussion, the concept of ``imbalance'' will serve as an important metric, and is defined in Equation \ref{e:1} for each processor in a given partition. 
\begin{equation}\label{e:1}
    \text{Imbalance}_i= \frac{|w_i-\sum ^N_{j=1}w_j/N|}{\sum ^N_{j=1}w_j} \times 100,
\end{equation}
where imbalance for the i-th processor is defined as the normalised difference between work allocated to that processor ($w_i$) and the ideal case of perfect distribution where each processor would have the same amount of work. The latter is found by simply dividing total work ($\sum ^N_{j=1}w_j$) by the number of desired processors ($N$). 

\begin{figure}[ht]
\centering
\begin{subfigure}[b]{0.49\linewidth}
\includegraphics[width=\linewidth]{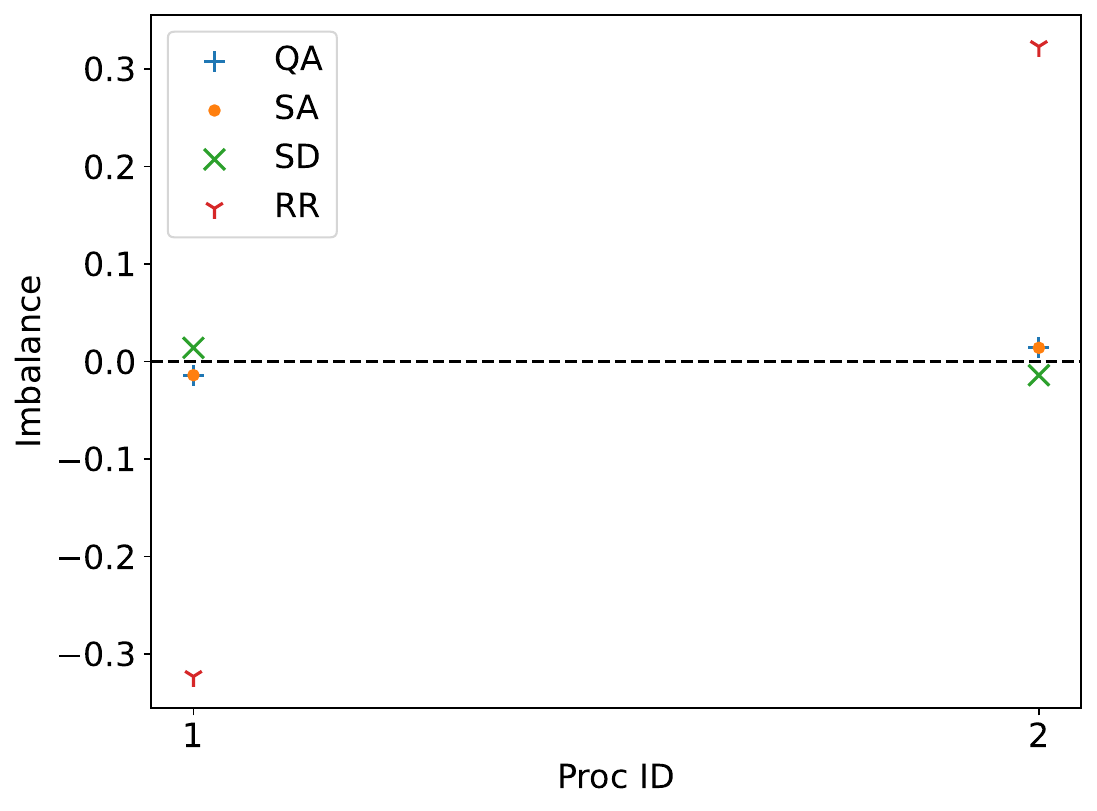}
\caption{Split across 2 processors}
\label{fig:2proc}
\end{subfigure}
\hfill
\begin{subfigure}[b]{0.49\linewidth}
\includegraphics[width=\linewidth]{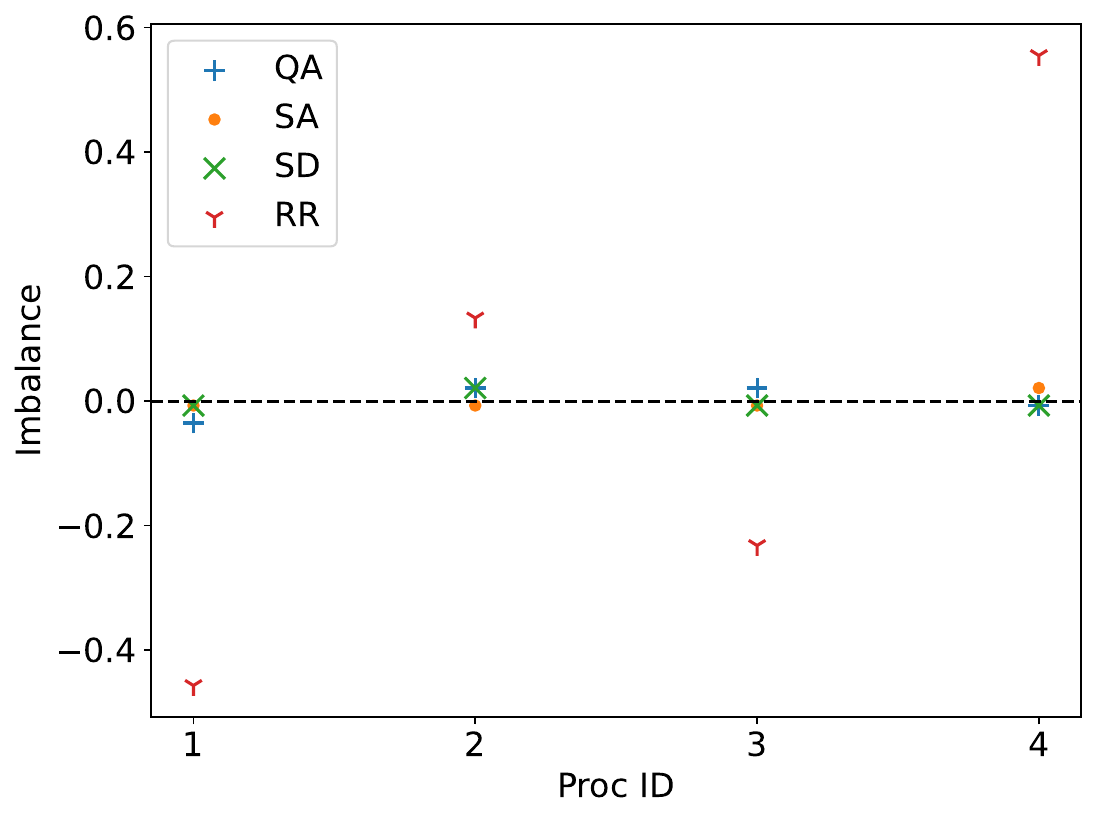}
\caption{Split across 4 processors}
\label{fig:4proc}
\end{subfigure}
\medskip
\begin{subfigure}[b]{0.49\linewidth}
\includegraphics[width=\linewidth]{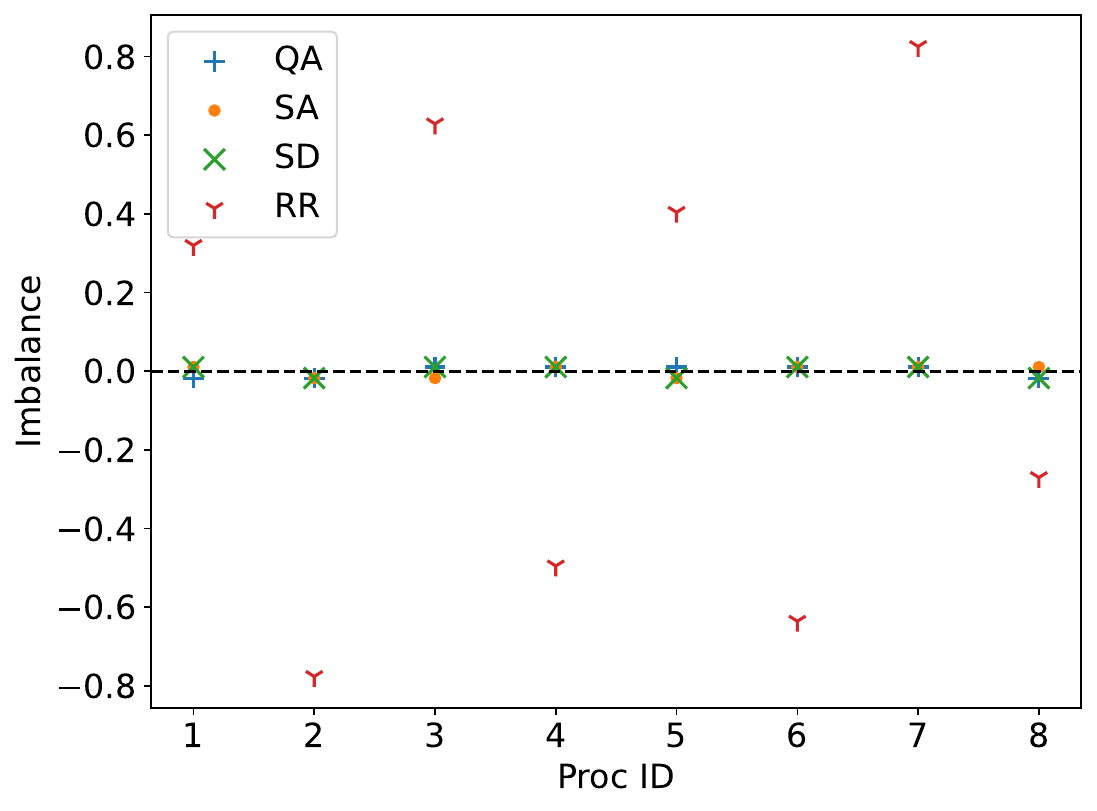}
\caption{Split across 8 processors}
\label{fig:8proc}
\end{subfigure}
\hfill
\begin{subfigure}[b]{0.49\linewidth}
\includegraphics[width=\linewidth]{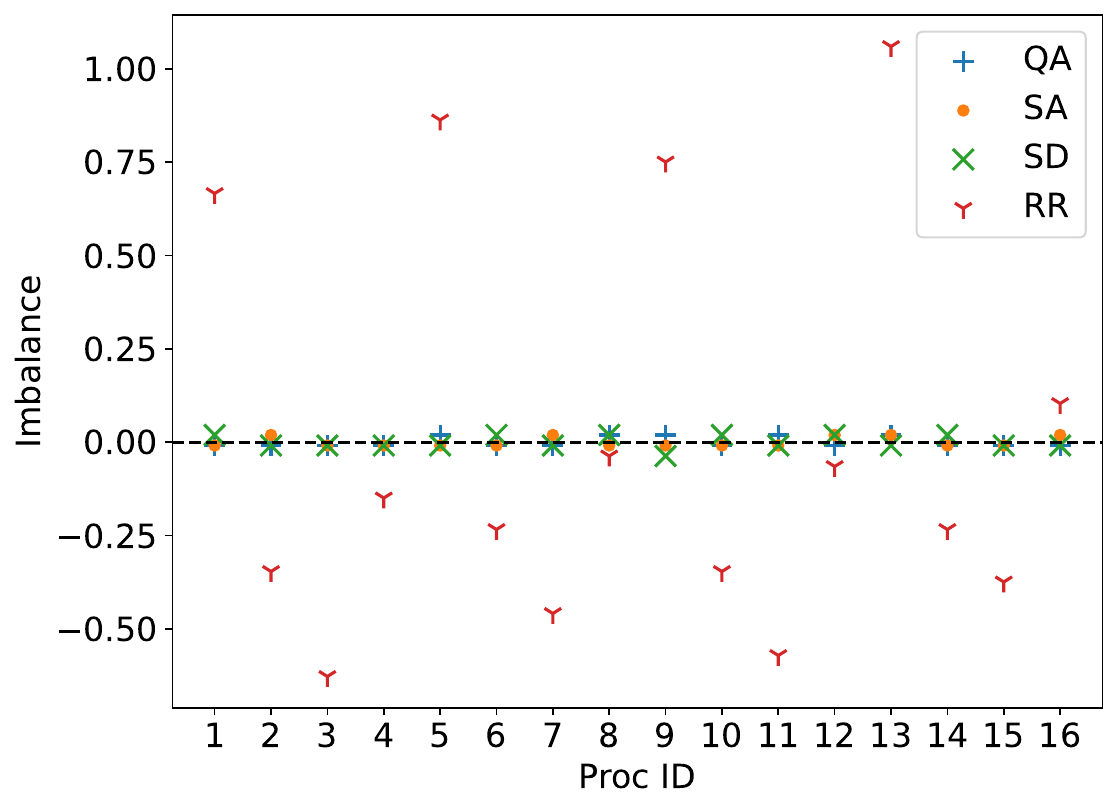}
\caption{Split across 16 processors}
\label{fig:16proc}
\end{subfigure}
\caption{Assigned imbalance for each processor after distributing total work between 2 (a), 4 (b), 8 (c) and 16 (d) cores. Partitioning was done via quantum annealing (QA), simulated annealing (SA), steepest descent (SD) and a round robin (RR) protocol. }
\label{fig:generalload}
\end{figure}

As an introductory example, Figure \ref{fig:generalload} illustrates the output from partitioning work from a simulation with approximately 100 grids across 2, 4, 8 and 16 processors. Essentially the problem reduces to number partitioning of 100 integers that is done recursively in order to obtain the desired number of subsets. Results from classical methods such as simulated annealing, steepest descent (SD) and a round robin (RR) strategy are included for comparison.
The quantum annealing consisted of 500 anneals with only the lowest energy solution displayed here. 
While a comprehensive analysis of parameter impact is reserved for later discussion, preliminary observations indicate some noteworthy trends. For instance, the QA-based approach consistently outperforms the Round Robin (RR) strategy and effectively circumvents local minima. The performance shows remarkable congruence with Simulated Annealing (SA), exhibiting negligible differences. A similar level of agreement is observed with the Steepest Descent (SD) method, based on the current dataset.

\begin{figure}
    \centering
    \includegraphics[width=0.5\linewidth]{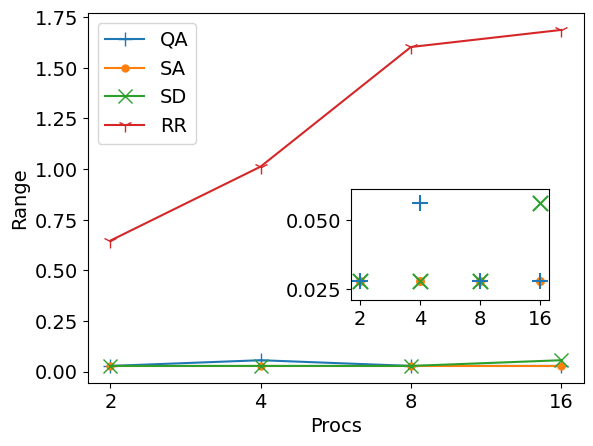}
    \caption{Range (i.e. maximum disparity in assigned work between processors) for partitions of 2, 4, 8 and 16 processors. This corresponds to the difference between subsets with the smallest and largest sums in the context of number partitioning. Solution strategies include quantum annealing (QA), simulated annealing (SA), steepest descent (SD) and round robin (RR). The inset provides a focus view without RR.}
    \label{fig:range}
\end{figure}

A more top-level performance metric to consider is the maximum disparity in processor load. This quantifies time spend by the faster processor waiting for the slowest one to complete its tasks before advancing to the next simulation time step. This \emph{range} between the fastest and slowest processors is graphically represented for the the same configurations as before in Figure \ref{fig:range}.
The round robin strategy emerges as the least efficient, leading to considerable CPU time wastage. In such a protocol, processors with fewer tasks remain idle for prolonged periods of time while awaiting the completion of assignments by processors with heavier loads. On the other hand, QA typically provides more balanced solutions. However, a distinct ``kink'' is observed when partitioning across four processors. This kink indicates a drop in solution quality compared to the results achieved with Simulated Annealing (SA) and Steepest Descent (SD).
This suboptimal performance can be attributed partially to the inherent stochastic nature of QA, where the optimal solution is not guaranteed in any specific run. However SA is also stochastic and performs better, so the QA solution may also be influenced by problem-specific parameters, a topic slated for in-depth evaluation in subsequent sections. Notably, when the computational work is distributed across a higher number of processors (e.g. 16), both QA and SA appear to surpass SD, which starts to falter due to the emergence of sub-optimal local minima in the increasingly fragmented energy landscape resulting from recursive partitioning. This suggests that, for some configurations, it may be possible to obtain an improvement even over more advanced classical methods such as SD.  

The aforementioned results pertain to a fixed number of anneals and a single invocation of the D-Wave API. While D-Wave functions in the minorminer library \cite{cai2014practical,zbinden2020embedding} effectively optimize the embedding of the logical problem onto the physical qubits, it is worthwhile noting that this algorithm is heuristic. A single call to the D-Wave API means that only one embedding configuration is utilized.
Under ideal conditions, this alone should not significantly impact the results. However, for the ensuing statistical analysis, five distinct embedding configurations were pre-computed and stored in order to map the problem to different physical qubits. 
When studying the effect of a parameter, for example number of anneals, each configuration underwent five separate runs using the pre-computed embeddings prior to averaging. 
This aims to minimize potential biases in the hardware that might undesirably affect the outcomes.

So far only the lowest energy solution was considered from the range of possible configurations that arise from QA. Despite demonstrating good performance this begs the important question of how likely is this to actually be observed in practice. Due to the inherent nature of load balancing requiring repeated redistribution after some number of time steps, it is not necessarily a stringent requirement to have the \emph{best} possible solution each time. In fact, it may suffice to have something close enough.

\begin{figure}[ht]
\centering
\begin{subfigure}[b]{0.49\linewidth}
    \includegraphics[width=\linewidth]{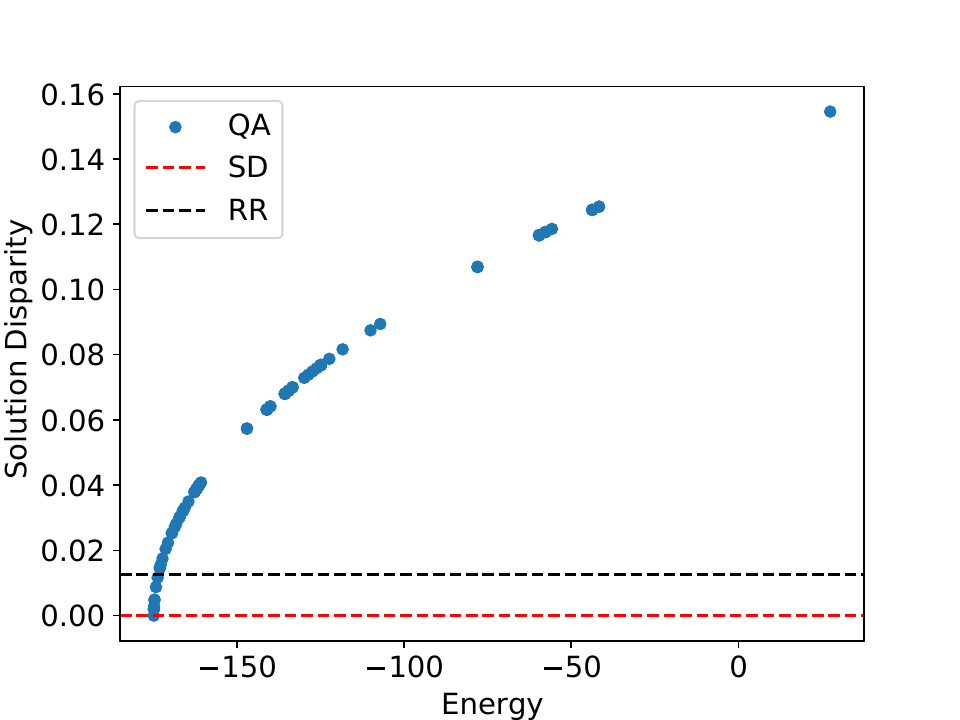}
    \caption{}
    \label{fig:sol_quality}
\end{subfigure}
\hfill
\begin{subfigure}[b]{0.49\linewidth}
\includegraphics[width=\linewidth]{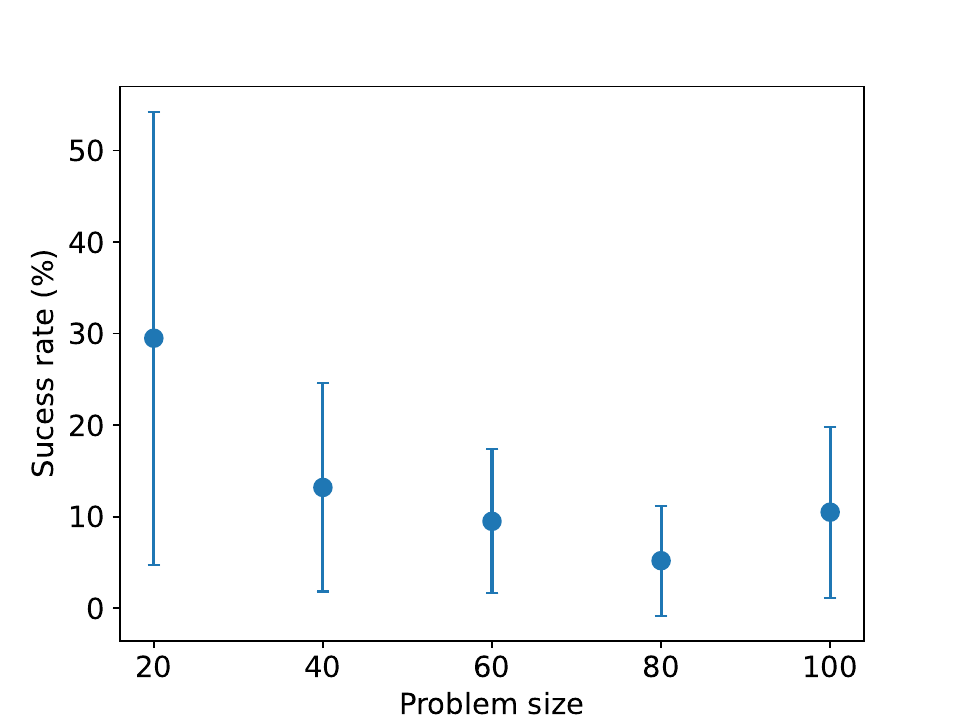}
\caption{}
\label{fig:succ}
\end{subfigure}
\caption{(a) Energy and solution disparity for all samples from a single call to the D-Wave API with 100 anneals for a problem size of 40 (QA), as well as solution disparity output from steepest descent (SD) and round robin (RR). A smaller solution disparity equates to a better solution quality. (b) Proportion of the 100 anneals that count as successful (i.e. better solution quality than RR) as a function of problem size.}
\end{figure}

Figure \ref{fig:sol_quality} attempts to illustrate this by plotting all the samples from a single run with a hundred anneals for a small problem size of 50 grids during a single partition. The energy output from a QA is arbitrarily scalable, therefore the actual numerical value is of little consequence other than the trend of a lower energies representing better solutions. This is in accordance with the corresponding solution disparity as defined in Equation \ref{e:2},
\begin{equation}\label{e:2}
    \text{Solution Disparity}=\frac{|w_1-w_2|}{0.5 \times (w_1+w_2)}.
\end{equation}
The solution disparity is thus the difference between assigned work loads for a single partition normalised by the work resulting from a perfect split. In other words, a lower solution disaparity corresponds with better solution quality. 

Included in Figure \ref{fig:sol_quality} for comparison are the solution disparities obtained from the deterministic round robin and steepest descent methods. Even with such a small number of anneals, the best (i.e. lowest energy) quantum solution has the same degree of imbalance as SD, with both methods arriving at the optimum solution for this configuration. There is also a significant proportion of the annealing samples that perform worse than SD but better than RR. These are the blue markers in between the two dotted, horizontal lines and suggest some degree of resilience in the method despite being heuristic. 
The spread of points with a higher disparity than the black dotted line are clearly inferior and hold little value. 

Figure \ref{fig:succ} illustrates the variation in the percentage of QA solutions that represent an improvement over RR across different problem sizes, based on results from 100 anneals. It is evident that the frequency of sub-optimal solutions escalates with the increase in problem size, indicating a relative under-performance compared to simulated annealing, which achieves a success rate of approximately $90\%$ across this range of problem sizes. Despite this, the quality of the most optimal solution obtained from both QA and SA remains the same.

This observation suggests that while QA has the potential to match the performance of more sophisticated classical algorithms, it does so for a limited fraction of the generated solutions. Given that QA is inherently a probabilistic method, it naturally involves generating a large number of samples to increase the likelihood of obtaining a high-quality solution. The advantage of QA lies in its ability to rapidly produce samples, with each annealing process taking just microseconds. As a result, even a relatively low success rate can, in practice, almost guarantee the identification of at least one successful outcome, which is often the primary goal. This efficiency in sample generation underscores the practical viability of QA.
Furthermore, it should be noted that at this point no \emph{a priori} optimisation for QA of user defined parameters such as number of anneals and chain strength have been provided to the annealer. Consequently, not only is the number of anneals very small here, but there is also a considerable number of chain breaks for problem sizes of 40 and more which can significantly degrade performance. 

\begin{figure}[h!]
\centering
\begin{subfigure}[b]{0.49\linewidth}
    \includegraphics[width=\linewidth]{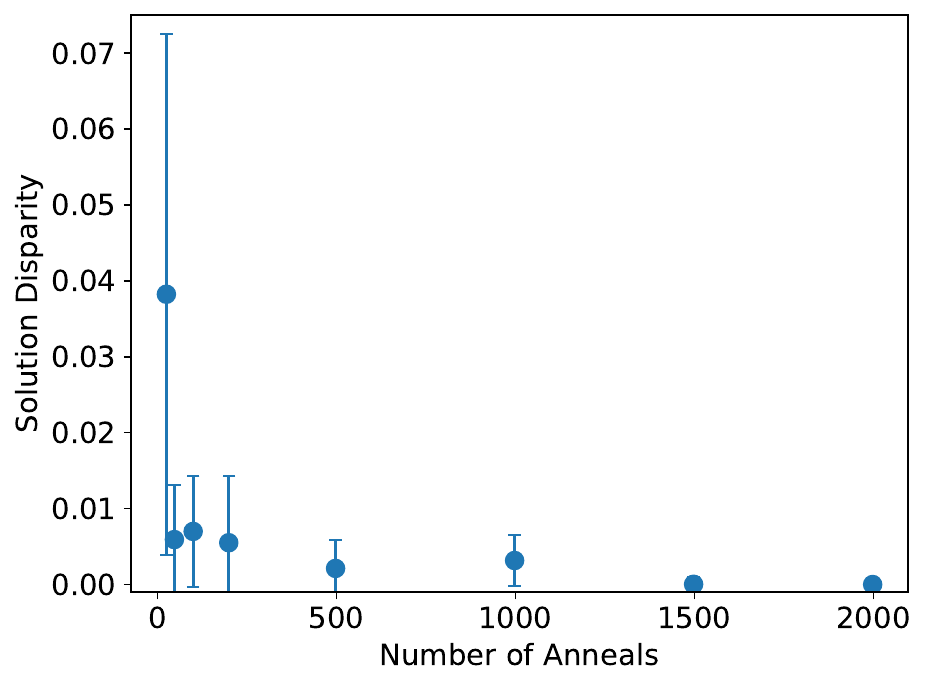}
    \caption{}
    \label{fig:n_anneals}
\end{subfigure}
\hfill
\begin{subfigure}[b]{0.49\linewidth}
\includegraphics[width=\linewidth]{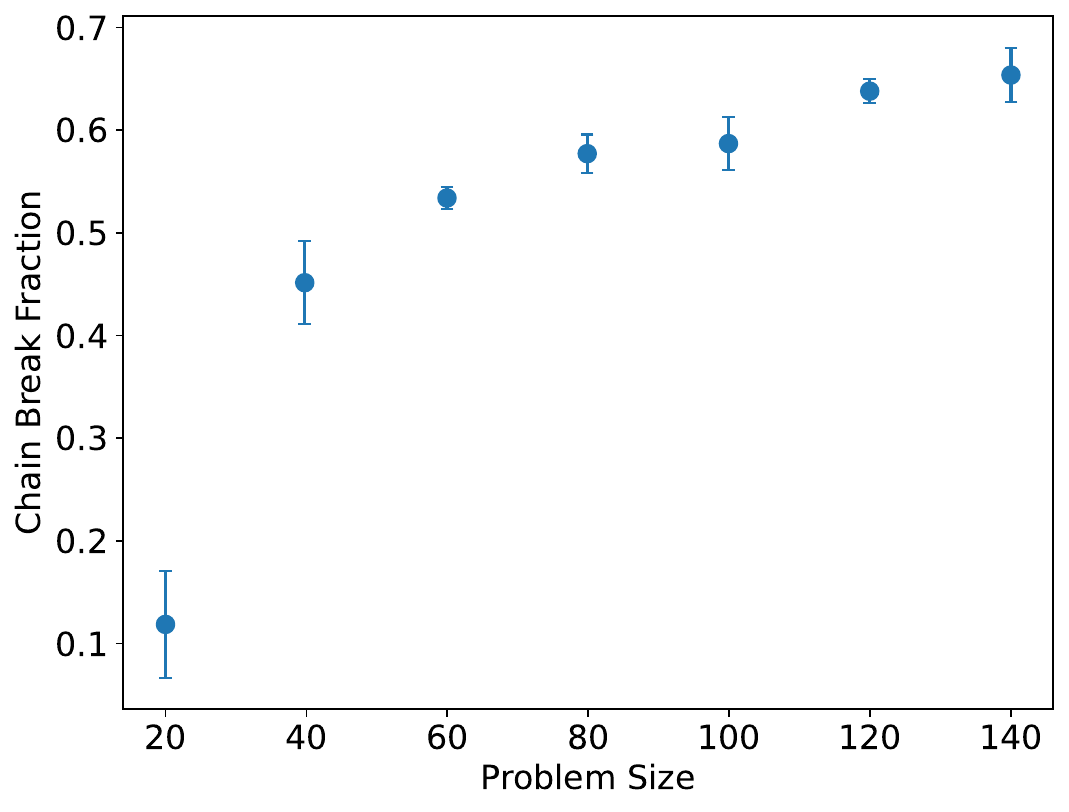}
\caption{}
\label{fig:cbf}
\end{subfigure}
\caption{(a) Impact of number of anneals on solution disparity for a fixed problem size of 100 grids. A lower solution disparity equates to a better quality solution, while negative values are not allowed by definition as per Equation \eqref{e:2}. (b) Mean chain break fraction as a function of increasing problem size (i.e. the number of grids being partitioned).}
\end{figure}

In order to evaluate the impact of parameter choices, the more challenging case with 100 grids is partitioned while changing the number of anneals. 
Figure \ref{fig:n_anneals} illustrates how the solution quality varies with anneals for a fixed problem size. Each configuration was repeated five times with a pre-computed embedding and the average lowest energy solution is shown here with error bars reflecting its standard deviation. 
As expected, increasing the number of anneals increases the likelihood of finding a near optimum solution. Although the mean quality seems to plateau relatively early in terms of magnitude, increasing anneals has significant impact on reducing error margins and thus the reliability of obtaining said solution. 
Considering the relatively small number of anneals needed to drastically improve solution quality as well as the cheap computation cost of each anneal, it is evident that this is likely not a limiting factor. 

Thus, the main roadblock to achieving scalability is unlikely to be the number of required anneals, but is perhaps the susceptibility to chain breaks. The Ising model here forms a fully connected graph while annealing hardware has limited physical couplings. This results in minor embedding forming chains of physical qubits to represent the same logical qubit. However it is evident in Figure \ref{fig:cbf} that the default chain strength, calculated using uniform torque compensation, quickly becomes inadequate at larger problem sizes.     
The significant increase in chain break fraction (CBF) as the number of patches/grids requiring partitioning increases result in a decision problem that is by default resolved via majority voting. At values as high as those seen in Figure \ref{fig:cbf} this has severe implications for the utility and robustness of the method for realistic applications where the number of patches/grids needing to be partitioned will be well over a hundred if not in the thousands or more. 

\begin{figure}[h!]
\centering
\begin{subfigure}[b]{0.49\linewidth}
    \includegraphics[width=\linewidth]{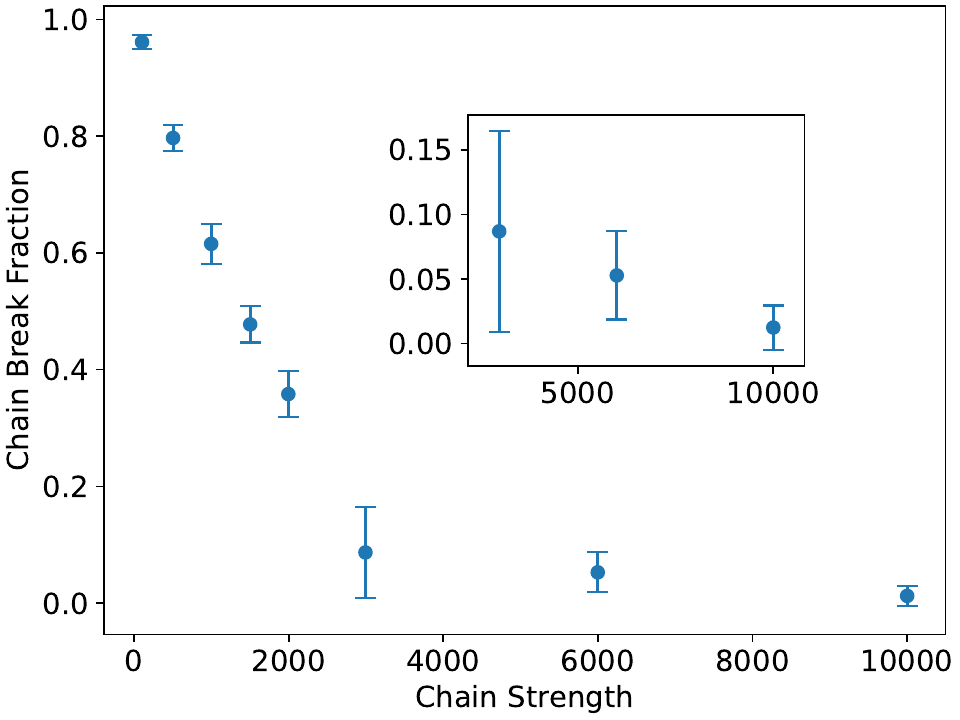}
    \caption{}
    \label{fig:cbf2}
\end{subfigure}
\hfill
\begin{subfigure}[b]{0.49\linewidth}
\includegraphics[width=\linewidth]{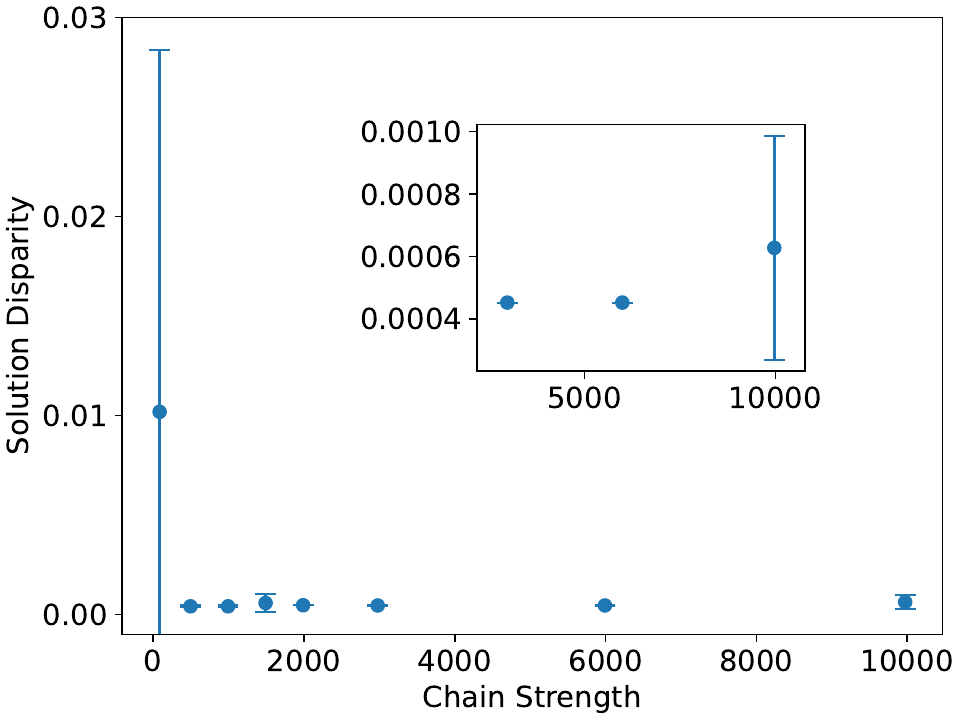}
\caption{}
\label{fig:cbf3}
\end{subfigure}
\caption{(a) Mean chain break fraction as a function of chain strength for a fixed problem size of 100 grids. (b) Mean solution disparity (i.e. lower disparity equates to a better solution) as a function of chain strength for the same 100 grid problem. Insets include a focus view on the rightmost three points.}
\end{figure}

One solution to this is overriding the default chain strength scheme and manually setting stronger chains. This approach was explored in Figure \ref{fig:cbf2} for a fixed problem size involving a hundred grids. The chain strength in the graph is defined by the multiplier applied to the max value in the data set. In other words, a chain strength of $1000$ here means a value a thousand times larger than the maximum number in the set. 
It is evident that the value required to maintain an acceptable fraction of breaks is several orders of magnitude higher than one might have anticipated from the size of numbers being partitioned. It is unclear if it would have been possible to predict this \emph{a priori} and how this trend would scale at realistic problem sizes.

It is also well known that blindly increasing the chain strength is not the perfect solution. Having qubits in very rigid chains can make them inflexible and less efficient at exploring the solution space. It can also wash out the other energy scales so that they are less than the device temperature and therefore lead to effectively random solutions.
Furthermore, the energy penalty from the chain term can start rivalling the energy contribution from the actual Ising model if it grows too large and this can introduce a bias in the solution. 
Thus there is a balancing act to be performed, where clearly the default scheme is not adequate for the problem being considered here, yet too strong a chain is also undesirable. This is demonstrated in Figure \ref{fig:cbf3} where increasing the chain strength brings a large improvement in solution quality. However, with very strong chains the solution does begin to degrade again. The silver lining is that there seems to be a very large region of the state space that results in good solutions.   

Taken together, the analysis here suggests that QA exhibits notable improvements when compared to basic classical algorithms like RR. Furthermore, QA demonstrates the capability to achieve solutions of comparable quality to those generated by more sophisticated algorithms, such as SA and SD. However, to attain such high-quality solutions consistently, QA may necessitate preliminary adjustments or tuning, highlighting the importance of optimizing the quantum annealing process for specific problem sets.
It is important to acknowledge that the scope of problem sizes investigated in this study is constrained by the limitations of current quantum computing hardware. As hardware capabilities improve, it is anticipated that these limitations will decrease, thereby expanding the range of problem sizes that can be efficiently addressed. 
The complexity of the energy landscape in this section is also limited, whereby advanced classical methods are able to arrive at near optimum solutions without being excessively hindered by local minima traps. We anticipate QA will be more advantageous over classical approaches for more complex energy landscapes with deep local minima traps. 
This highlights a crucial aspect of quantum annealing: identifying and formulating problem sets that truly leverage the unique strengths of quantum algorithms to solve problems that are intractable for classical methods. 
\subsection*{Particle Based Application}
\begin{figure}[h!]
\centering
\begin{subfigure}[b]{0.49\linewidth}
    \includegraphics[width=\linewidth]{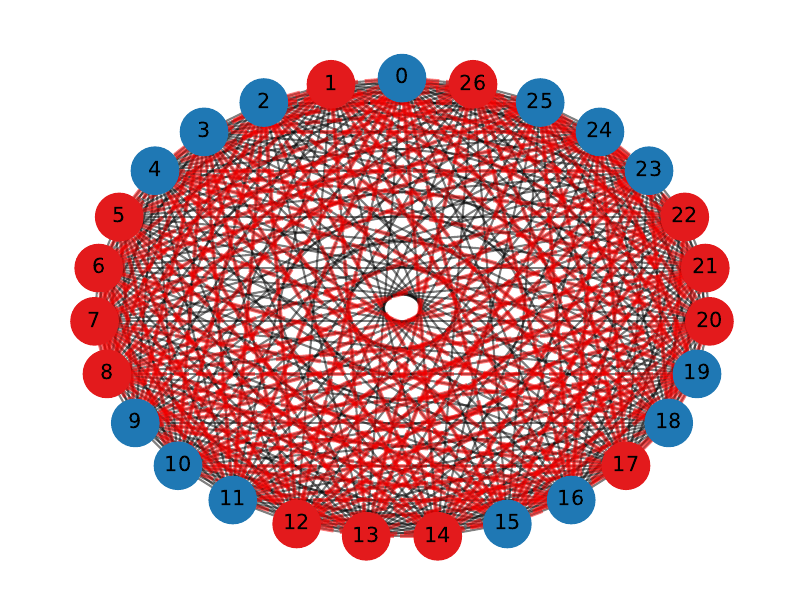}
    \caption{}
    \label{fig:gen2}
\end{subfigure}
\hfill
\begin{subfigure}[b]{0.49\linewidth}
\includegraphics[width=\linewidth]{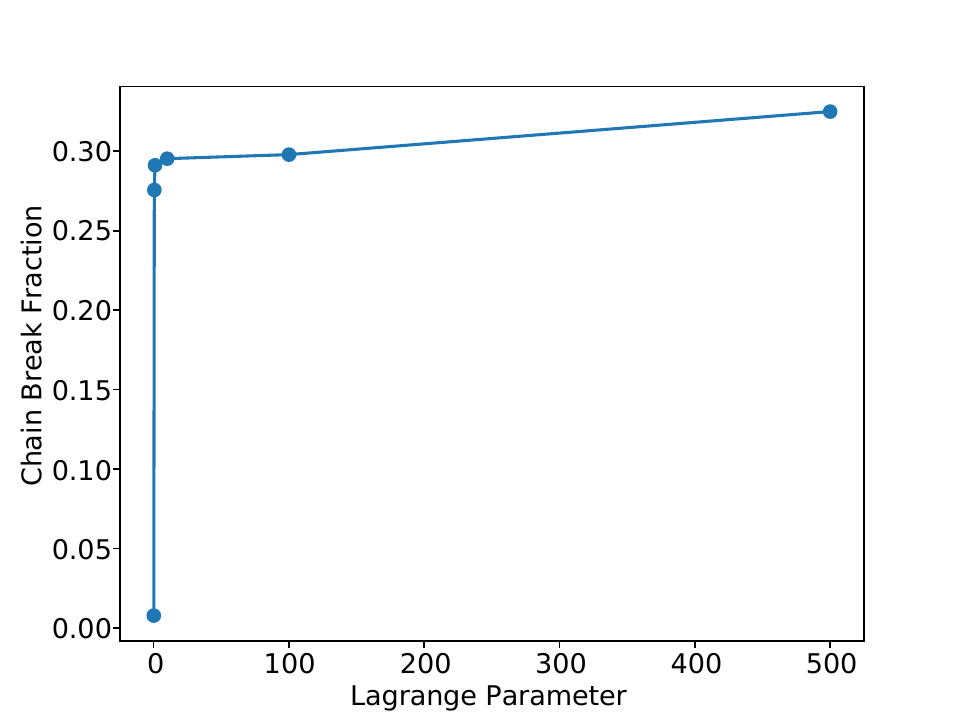}
\caption{}
\label{fig:cbf4}
\end{subfigure}
\caption{(a) The 27 node, fully connected load balancing graph for SWIFT. Blue and red circles represent nodes in different subsets and red/black lines represent cut/uncut edges following a partition.(b) Change in chain break fraction as a function of Lagrange parameter. The Lagrange parameter determines the relative importance between penalty terms in the Ising model. A larger value places more significance on minimising the difference in node weights while a smaller value signifies greater value in minimising cut edge weights.}
\end{figure}

The graph representing load balancing for SPH is incredibly dense as illustrated in Figure \ref{fig:gen2} due to its fully connected nature. An optimised partition should aim to minimise the mismatch between subsets of total node weight, while also minimising the sum of cut edge weights. It is not a clear \emph{a priori} which of the two objectives is more important as this is likely to be somewhat dependant on the HPC architecture, in particular whether intra or inter processor bandwidth is the more limiting factor for the CPU stack. For example, in the case of the latter, it would be more strategic to further minimise cut edge weight where possible even at the cost of a slightly higher node imbalance and vice versa for the former. 
Therefore in order to remain general, the task will be considered a multi-objective optimisation problem here.   

The Lagrange parameter, $\gamma$, determines the relative importance between these two objectives as shown in Equation \ref{e99}. A high value implies more significance attributed to minimising the difference of node weights between the two subsets. In the limit of very large $\gamma$ the problem essentially reduces back to number partitioning since the edges will have negligible relative influence. An interesting finding was that changing this parameter had an unexpected impact on the chain break fraction. 
This is shown in Figure \ref{fig:cbf4} where a small value results in lower chain breaks than was observed in the grid based, number partitioning example even for a graph of the same size. Indeed, as the Lagrange parameter is increased, the problem tends back towards number partitioning and chain breaks become more frequent.
This is despite no change to the underlying graph structure, which remains a $27$ node clique. Moreover, numerical values are auto-scaled in the process of mapping onto a quantum processor and so differences in the values of weights alone should not be significantly impacting chain breaks. 
This implies that despite not changing the nature of the underlying embedding, the load balancing problem here is intrinsically more resilient to chain breaks than for its grid based counterpart.   

\begin{figure*}[ht]
\centering
\begin{subfigure}{0.5\textwidth}
\includegraphics[width=\linewidth]{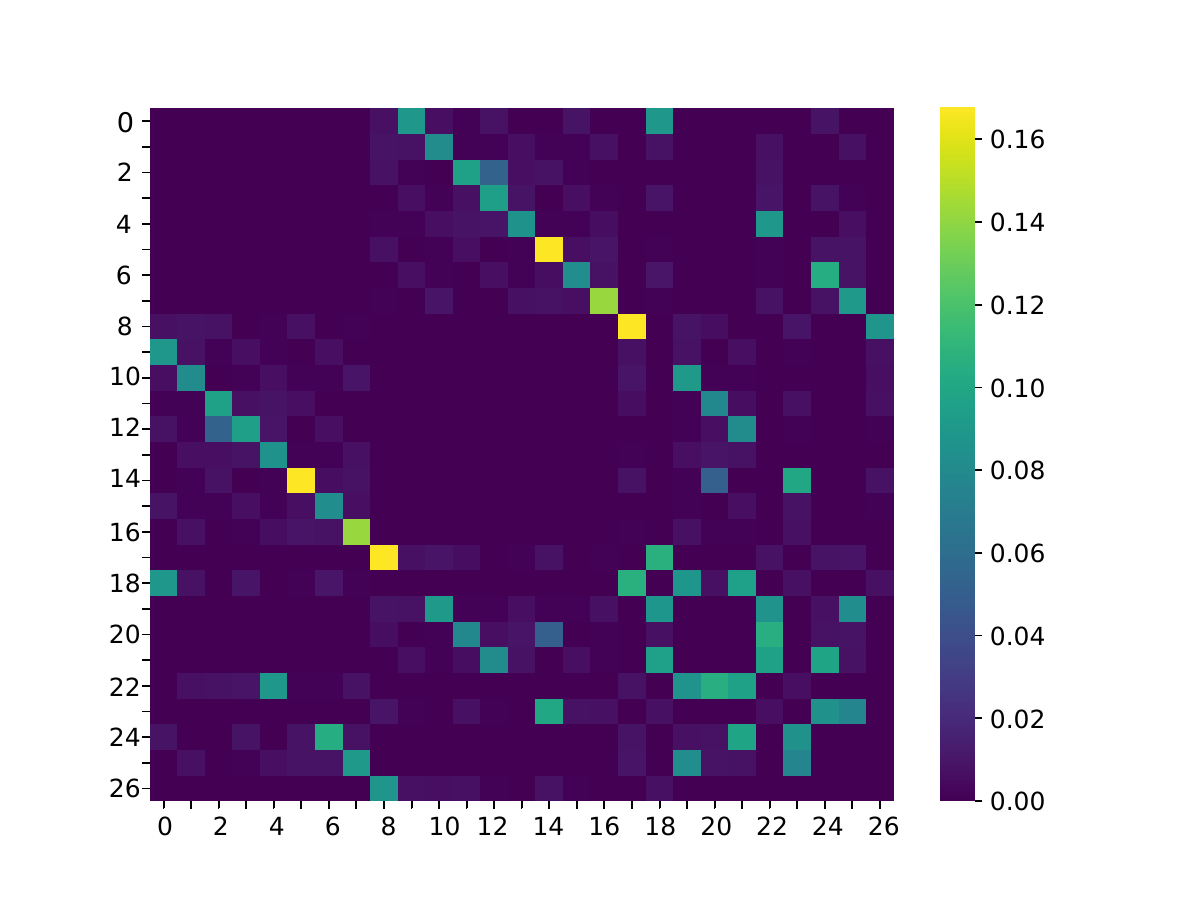}
\caption{QA}
\label{fig:sub1}
\end{subfigure}%
\hspace{-10mm} 
\begin{subfigure}{0.5\textwidth}
\includegraphics[width=\linewidth]{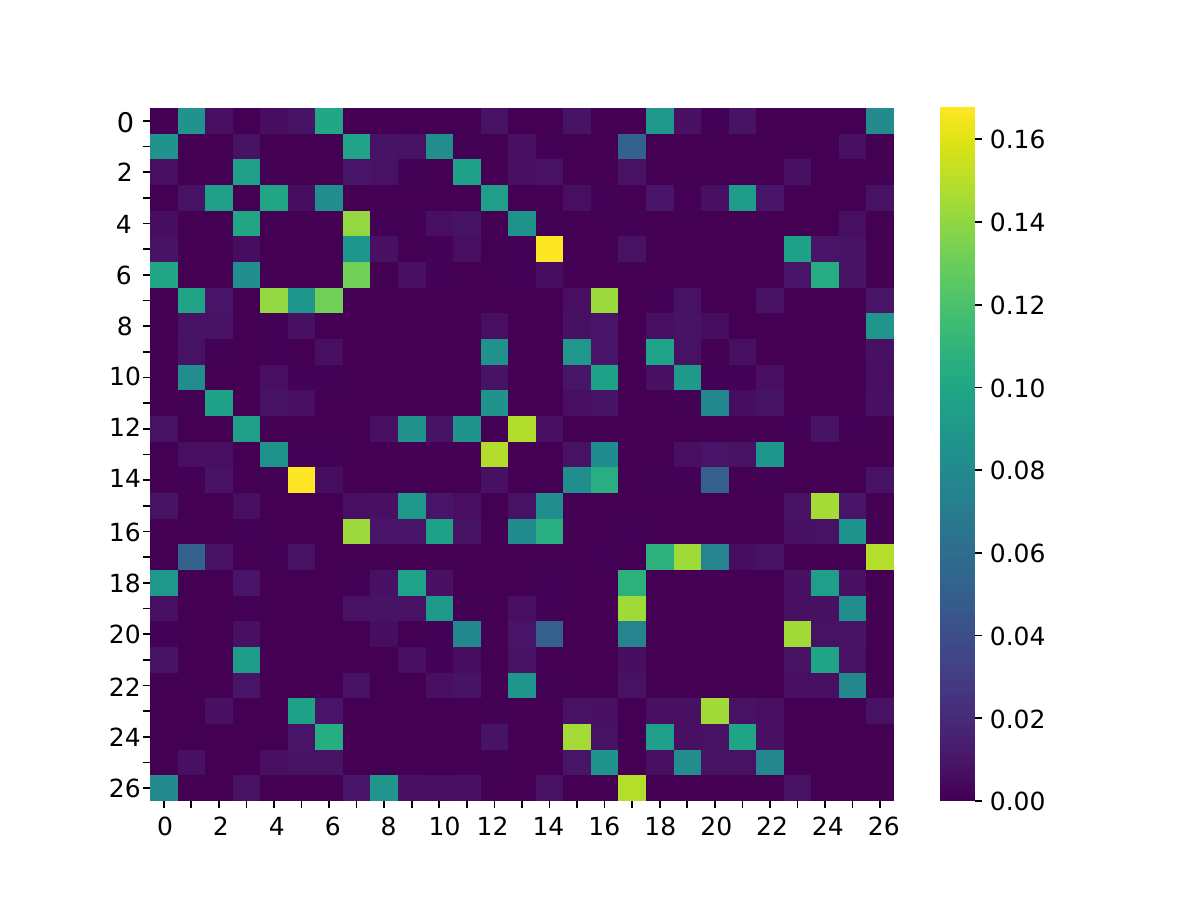}
\caption{METIS}
\label{fig:sub2}
\end{subfigure}
\caption{Heat maps for the partitions from QA and METIS respectively. Each entry in the matrix represents an edge between two nodes with the corresponding indices along the outer edges. The color intensity indicates the normalised weight of a cut edge. A higher proportion of low weight entries (i.e. light purple) implies a better partition as opposed to more highly weighted cut edges (i.e. green/yellow).}
\label{fig:heat}
\end{figure*}

Consider for now a neutral value for the Lagrange parameter of unity. 
A single QA was conducted with one thousand anneals and the lowest energy solution compared to a partition obtained using METIS, a state of the art classical graph partitioning software. 
This is illustrated in Figure \ref{fig:heat}, which displays the cut edge matrices as heat maps. 
Each entry in the matrix represents the edge connecting nodes with the corresponding axis indices, while the color intensity is indicative of the weight of said edge. A color value of null (i.e. dark purple) indicates that edge was not cut. Both QA and METIS make an exact total of $182$ cuts each and share the same magnitude in terms of single largest cut. 
However, QA consistently makes better choices in exploring the energy landscape and manages to sever less expensive edges for a combined saving of close to $33\%$. The combined weight of cut edges from this QA run was only $66\%$ the size of its METIS counterpart. Furthermore, this came with a better node balance as well where the degree of imbalance using QA was only $34\%$ the size of imbalance allocated by METIS.

Repeating the QA runs five times and averaging, the same trend still stands as indicated in Table \ref{table:1}, suggesting a resilience of the method to probabilistic fluctuations. Note that node and edge weights have been normalised by the same factor for both methods as they operate on the same data set. Furthermore the ``Performance Ratio'' entries in the table are simply the fractional result of dividing the corresponding METIS entry by the QA counterpart. As both objectives are tailored to be minimised, a larger than unity performance ratio indicates some quantum advantage. Moreover QA performs better across both objectives simultaneously.   

\begin{table}[ht]
\caption{Average performance statistics for quantum annealing compared with METIS. The ideal solution has as small a value as possible for both solution imbalance and cut edge weights.}
\centering
\label{table:1}
\begin{tabular}{lcc}  
\toprule
& Solution Disparity & Cut Edge Weights \\
\midrule
Quantum Annealing & 0.057 & 3.69 \\
METIS & 0.189 & 5.20 \\
Performance Ratio & 3.32 & 1.41 \\  
\bottomrule
\end{tabular}
\end{table}

\begin{figure}[h!]
    \centering
    \includegraphics[width=0.5\linewidth]{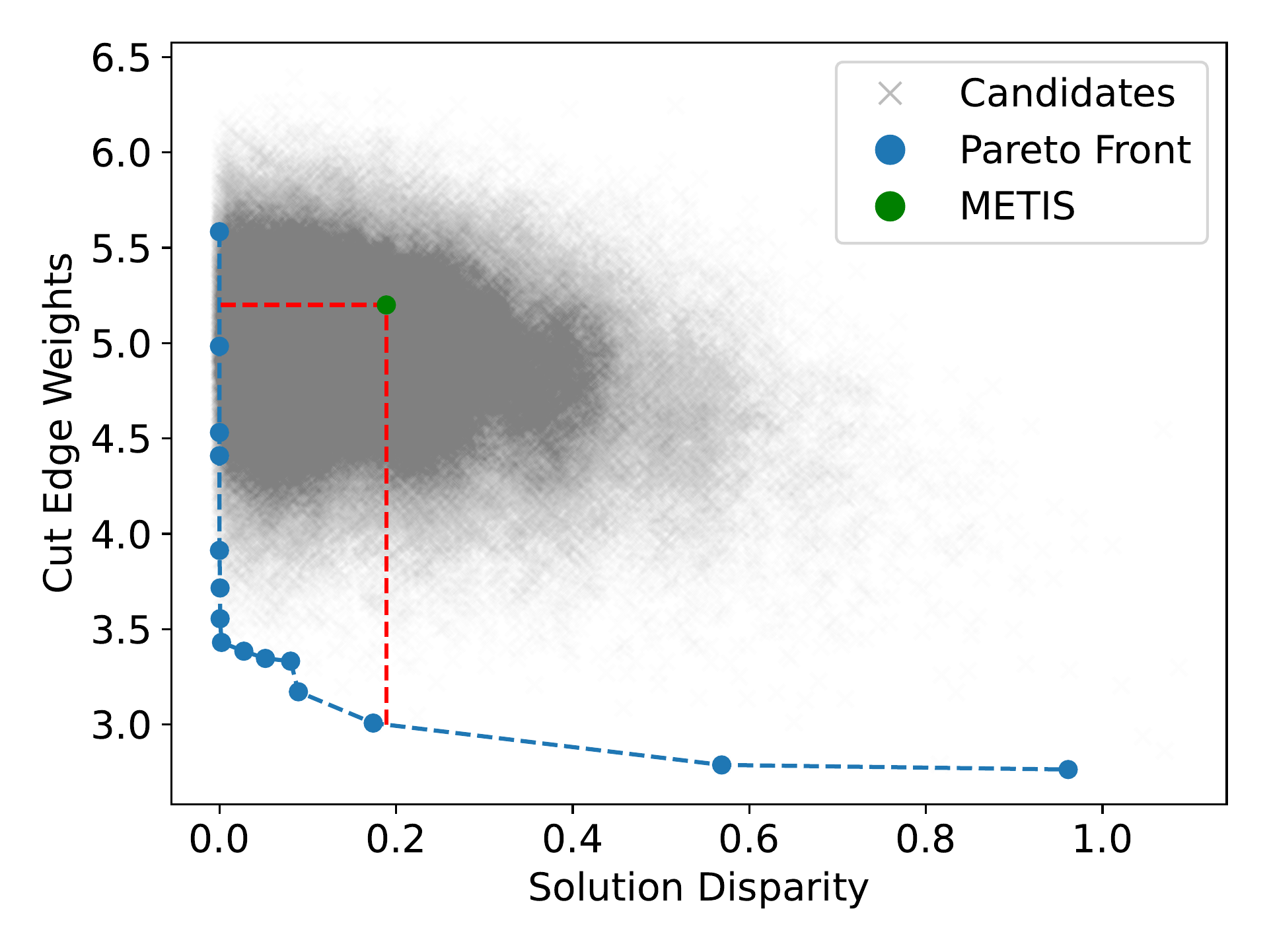}
    \caption{Approximate Pareto front as mapped out by QA for optimising (i.e. minimising) cut edge weight and solution disparity. Included are all the other solutions obtained by QA as well as the output from METIS. A significant proportion (close to $41\%$) of QA samples that are sub-optimal to the Pareto front, are still Pareto dominant  when compared to METIS. These are the points to the left and below the METIS solution as indicated by the area encapsulated by red dashed lines and the Pareto front.}
    \label{fig:pareto}
\end{figure}

In addition to this, the outcomes of quantum annealing can be further fine-tuned to meet the specific needs of the HPC cluster by adjusting the Lagrange parameter. To maintain a broad applicability, this study extensively explored the parameter's state space through 100 iterations with 1000 anneals each and evenly spaced values for the Lagrange parameter between 0 and 50. The Pareto dominant solutions in terms of the two objective functions obtained from this process are presented in Figure \ref{fig:pareto}. The Pareto front here represents a set of solutions for which one objective function cannot be improved without bringing about a detriment to the second objective. The right side of the Pareto front is fairly close to horizontal, suggesting that if one approaches from the far right, vast improvements can be made in terms of solution disparity at the cost of a very minimal increase to cut edge weights. Conversely, initially starting at the left and proceeding towards the right large improvements to the cut edges are attainable at little cost to the node balance. The solution from METIS is also included for comparison and evidently inferior in that it is clearly possible to improve both objectives. Moreover the latter is true not just for a handful of lucky anneals, but for a large proportion as observed with the collection of QA points that lie off of the Pareto front but still are Pareto dominant in relation to the METIS output. This corresponds to close to 41000 of the total 100,000 sample solutions and demonstrates the resiliency of QA for complex problems. 
Even outside this region, most candidate solutions are only inferior to METIS in one objective, likely as a result of enforcing either very large or very small Lagrange parameters. This will naturally prioritise one objective even at the potential detriment of the other. 

\newpage
\section*{Conclusions and Outlook}
This study explored the utilization of quantum annealing as a strategic approach for the critical task of workload allocation to processors in parallel HPC applications. Initially focusing on less complex, grid-based applications, QA demonstrated improvements over simpler classical strategies, yet it did not conclusively surpass more sophisticated classical methods. A significant challenge for QA, unlike these classical methods, is the limited connectivity between physical qubits. This limitation results in a critical barrier to scalability, particularly given the fully connected nature of the problem at hand. 

However, this drawback is somewhat mitigated in the context of the second application, which focused on load balancing in particle based codes. While the smallest problem configuration which was studied here is fully connected due to each cell's proximity within the system, this is not expected to hold for larger problems. Recall that SPH kernels are characterized by compact support, this implies that larger problem graphs, while more extensive, are not likely to be fully connected. This subtlety enhances the feasibility of applying QA, especially in the Noisy Intermediate Scale Quantum (NISQ) era where inter-qubit connectivity can sometimes be a more critical constraint than the total number of qubits. 
Moreover, for simpler 2D/quasi-planar configurations this might not be as severe a constraint even for larger problem sizes. This is due to the compatibility between the problem graph and alignment of qubits with their couplers, which when implemented on a chip are inherently quasi-planar as well.   

In addition, QA demonstrated an improvement in performance even against a state of the art classical method, despite the problem's evolution into a complex multi-objective optimization. The Lagrange parameter formulation allows the user to explore the Pareto front and obtain highly optimised solutions tailored towards individual hardware architectures.  Although this necessitated running a representative simulation and partitioning it repeatedly to explore the state space here, in general more cost effective methods such as machine learning could be used to estimate a good value \emph{a priori}. This approach draws strong parallels with existing efforts that have endeavoured to do so for some of the other annealing parameters\cite{barbosa2021optimizing,barbosa2021using}. The Ising formulation can also be readily extended to allow concurrent partitioning into multiple subsets simultaneously rather than recursively \cite{ushijima2017graph} to better accommodate larger problems. 

Given the observed enhancements in solution quality and the reduced connectivity demands in larger problem graphs, it is conceivable that as annealers continue to evolve, they may become viable for graph partitioning tasks such as this. 
This is particularly relevant for hybrid applications such as load balancing where the majority of the algorithm could still be executed on CPUs, with only a select, complex segment offloaded to the quantum processor in tandem. As such, perhaps a heterogeneous architecture integrating both quantum and classical computing resources could be a strategic path forward. Particularly so in light of recent endeavours to strive towards such integrated systems \cite{cavallaro2022hybrid,lippert2022perspectives}. Moreover, one of the most time consuming components of the QA algorithm is currently the communication time between classical and quantum hardware and associated latency costs. This would be crucially reduced to almost negligible levels by co-locating the respective chips in the same data center. The time for each anneal itself remains competitive at around $20\mu s$ compared to the roughly $15\mu s$ needed by METIS and is likely to improve with evolving hardware. However, this will be of comparatively little consequence as co-location will allow the system to run in parallel, thus best leveraging the inherent strengths of each respective system.

\section*{Acknowledgements}
The authors would like to thank Dr Peter Draper at Durham University and Dr Matthieu Schaller at Leiden University for insightful discussions. Similarly, the authors are grateful to Prof Vivien Kendon at University of Strathclyde for useful comments and discussion on this work, as well as the rest of the QEVEC/QuANDiE teams. The authors acknowledge funding through UKRI EPSRC projects (EP/W00772X/2 and EP/Y004515/1).
This work used the DiRAC@Durham facility managed by the Institute for Computational Cosmology on behalf of the STFC DiRAC HPC Facility (www.dirac.ac.uk). The equipment was funded by BEIS capital funding via STFC capital grants ST/K00042X/1, ST/P002293/1, ST/R002371/1 and ST/S002502/1, Durham University and STFC operations grant ST/R000832/1. DiRAC is part of the National e-Infrastructure.
\bibliography{sample}

\begin{thebibliography}{10}

\bibitem{danowitz2012cpu}
Andrew Danowitz, Kyle Kelley, James Mao, John~P Stevenson, and Mark Horowitz.
\newblock Cpu db: Recording microprocessor history: With this open database,
  you can mine microprocessor trends over the past 40 years.
\newblock {\em Queue}, 10(4):10--27, 2012.

\bibitem{sutter2005free}
Herb Sutter et~al.
\newblock The free lunch is over: A fundamental turn toward concurrency in
  software.
\newblock {\em Dr. Dobb’s journal}, 30(3):202--210, 2005.

\bibitem{tan2009influence}
Weiliang Tan, Peizhen Lin, Bo~Liang, and Hui Deng.
\newblock Influence of network bandwidth on parallel computing performance with
  intra-node and inter-node communication.
\newblock In {\em 2009 Second International Conference on Intelligent Networks
  and Intelligent Systems}, pages 534--537. IEEE, 2009.

\bibitem{rabenseifner2003hybrid}
Rolf Rabenseifner et~al.
\newblock Hybrid parallel programming on hpc platforms.
\newblock In {\em proceedings of the Fifth European Workshop on OpenMP, EWOMP},
  volume~3, pages 185--194, 2003.

\bibitem{schwarz1869ueber}
Hermann~Amandus Schwarz.
\newblock {\em Ueber einige Abbildungsaufgaben: aus einer Mittheilung an Herrn
  Richelot in K{\"o}nigsberg}.
\newblock 1869.

\bibitem{hessenius1982zonal}
K~Hessenius and T~Pulliam.
\newblock A zonal approach to solution of the euler equations.
\newblock In {\em 3rd Joint Thermophysics, Fluids, Plasma and Heat Transfer
  Conference}, page 969, 1982.

\bibitem{henshaw1987multigrid}
William~D Henshaw and G~Chesshire.
\newblock Multigrid on composite meshes.
\newblock {\em SIAM Journal on Scientific and Statistical Computing},
  8(6):914--923, 1987.

\bibitem{glowinski1983domain}
R~Glowinski, QV~Dinh, and J~Periaux.
\newblock Domain decomposition methods for nonlinear problems in fluid
  dynamics.
\newblock {\em Computer methods in applied mechanics and engineering},
  40(1):27--109, 1983.

\bibitem{keyes2013multiphysics}
David~E Keyes, Lois~C McInnes, Carol Woodward, William Gropp, Eric Myra,
  Michael Pernice, John Bell, Jed Brown, Alain Clo, Jeffrey Connors, et~al.
\newblock Multiphysics simulations: Challenges and opportunities.
\newblock {\em The International Journal of High Performance Computing
  Applications}, 27(1):4--83, 2013.

\bibitem{houzeaux2017domain}
G~Houzeaux, JC~Cajas, Marco Discacciati, B~Eguzkitza, A~Gargallo-Peir{\'o},
  M~Rivero, and M~V{\'a}zquez.
\newblock Domain decomposition methods for domain composition purpose: Chimera,
  overset, gluing and sliding mesh methods.
\newblock {\em Archives of Computational Methods in Engineering},
  24:1033--1070, 2017.

\bibitem{tang2021review}
HS~Tang, RD~Haynes, and G~Houzeaux.
\newblock A review of domain decomposition methods for simulation of fluid
  flows: Concepts, algorithms, and applications.
\newblock {\em Archives of Computational Methods in Engineering}, 28:841--873,
  2021.

\bibitem{wang2010adaptive}
Peng Wang, Tom Abel, and Ralf Kaehler.
\newblock Adaptive mesh fluid simulations on gpu.
\newblock {\em New Astronomy}, 15(7):581--589, 2010.

\bibitem{bennett1985structural}
JA~Bennett and ME~Botkin.
\newblock Structural shape optimization with geometric description and adaptive
  mesh refinement.
\newblock {\em AIAA journal}, 23(3):458--464, 1985.

\bibitem{botti2010adaptive}
Lorenzo Botti, Marina Piccinelli, Bogdan Ene-Iordache, Andrea Remuzzi, and Luca
  Antiga.
\newblock An adaptive mesh refinement solver for large-scale simulation of
  biological flows.
\newblock {\em International Journal for Numerical Methods in Biomedical
  Engineering}, 26(1):86--100, 2010.

\bibitem{zumbusch2012parallel}
Gerhard Zumbusch.
\newblock {\em Parallel multilevel methods: adaptive mesh refinement and
  loadbalancing}.
\newblock Springer Science \& Business Media, 2012.

\bibitem{dubey2014survey}
Anshu Dubey, Ann Almgren, John Bell, Martin Berzins, Steve Brandt, Greg Bryan,
  Phillip Colella, Daniel Graves, Michael Lijewski, Frank L{\"o}ffler, et~al.
\newblock A survey of high level frameworks in block-structured adaptive mesh
  refinement packages.
\newblock {\em Journal of Parallel and Distributed Computing},
  74(12):3217--3227, 2014.

\bibitem{gingold1977smoothed}
Robert~A Gingold and Joseph~J Monaghan.
\newblock Smoothed particle hydrodynamics: theory and application to
  non-spherical stars.
\newblock {\em Monthly notices of the royal astronomical society},
  181(3):375--389, 1977.

\bibitem{monaghan2012smoothed}
Joseph~J Monaghan.
\newblock Smoothed particle hydrodynamics and its diverse applications.
\newblock {\em Annual Review of Fluid Mechanics}, 44:323--346, 2012.

\bibitem{shadloo2016smoothed}
M~Safdari Shadloo, G~Oger, and David Le~Touz{\'e}.
\newblock Smoothed particle hydrodynamics method for fluid flows, towards
  industrial applications: Motivations, current state, and challenges.
\newblock {\em Computers \& Fluids}, 136:11--34, 2016.

\bibitem{lind2020review}
Steven~J Lind, Benedict~D Rogers, and Peter~K Stansby.
\newblock Review of smoothed particle hydrodynamics: towards converged
  lagrangian flow modelling.
\newblock {\em Proceedings of the royal society A}, 476(2241):20190801, 2020.

\bibitem{kadowaki1998quantum}
Tadashi Kadowaki and Hidetoshi Nishimori.
\newblock Quantum annealing in the transverse ising model.
\newblock {\em Physical Review E}, 58(5):5355, 1998.

\bibitem{lucas2014ising}
Andrew Lucas.
\newblock Ising formulations of many np problems.
\newblock {\em Frontiers in physics}, 2:5, 2014.

\bibitem{camino2023quantum}
B~Camino, J~Buckeridge, PA~Warburton, V~Kendon, and SM~Woodley.
\newblock Quantum computing and materials science: A practical guide to
  applying quantum annealing to the configurational analysis of materials.
\newblock {\em Journal of Applied Physics}, 133(22), 2023.

\bibitem{ushijima2017graph}
Hayato Ushijima-Mwesigwa, Christian~FA Negre, and Susan~M Mniszewski.
\newblock Graph partitioning using quantum annealing on the d-wave system.
\newblock In {\em Proceedings of the Second International Workshop on Post
  Moores Era Supercomputing}, pages 22--29, 2017.

\bibitem{collins2022ibm}
Hugh Collins and Chris Nay.
\newblock Ibm unveils 400 qubit-plus quantum processor and next-generation ibm
  quantum system two, 2022.

\bibitem{pressrelease2023}
D-Wave~Quantum Inc.
\newblock D-wave announces 1,200+ qubit advantage2™ prototype in new,
  lower-noise fabrication stack, demonstrating 20x faster time-to-solution on
  important class of hard optimization problems, 2024.
\newblock [Press release].

\bibitem{katzgraber2014glassy}
Helmut~G Katzgraber, Firas Hamze, and Ruben~S Andrist.
\newblock Glassy chimeras could be blind to quantum speedup: Designing better
  benchmarks for quantum annealing machines.
\newblock {\em Physical Review X}, 4(2):021008, 2014.

\bibitem{katzgraber2015seeking}
Helmut~G Katzgraber, Firas Hamze, Zheng Zhu, Andrew~J Ochoa, and Humberto
  Munoz-Bauza.
\newblock Seeking quantum speedup through spin glasses: The good, the bad, and
  the ugly.
\newblock {\em Physical Review X}, 5(3):031026, 2015.

\bibitem{boixo2016computational}
Sergio Boixo, Vadim~N Smelyanskiy, Alireza Shabani, Sergei~V Isakov, Mark
  Dykman, Vasil~S Denchev, Mohammad~H Amin, Anatoly~Yu Smirnov, Masoud Mohseni,
  and Hartmut Neven.
\newblock Computational multiqubit tunnelling in programmable quantum
  annealers.
\newblock {\em Nature communications}, 7(1):10327, 2016.

\bibitem{denchev2016computational}
Vasil~S Denchev, Sergio Boixo, Sergei~V Isakov, Nan Ding, Ryan Babbush, Vadim
  Smelyanskiy, John Martinis, and Hartmut Neven.
\newblock What is the computational value of finite-range tunneling?
\newblock {\em Physical Review X}, 6(3):031015, 2016.

\bibitem{yarkoni2022quantum}
Sheir Yarkoni, Elena Raponi, Thomas B{\"a}ck, and Sebastian Schmitt.
\newblock Quantum annealing for industry applications: Introduction and review.
\newblock {\em Reports on Progress in Physics}, 2022.

\bibitem{bauza2024scaling}
Humberto~Munoz Bauza and Daniel~A Lidar.
\newblock Scaling advantage in approximate optimization with quantum annealing.
\newblock {\em arXiv preprint arXiv:2401.07184}, 2024.

\bibitem{chancellor2020toward}
Nicholas Chancellor, Robert Cumming, and Tim Thomas.
\newblock Toward a standardized methodology for constructing quantum computing
  use cases.
\newblock {\em arXiv preprint arXiv:2006.05846}, 2020.

\bibitem{mena2022transparent}
J~Aguilar Mena, Omar Shaaban, Victor Lopez, Marta Garcia, Paul Carpenter,
  E~Ayguad, and Jesus Labarta.
\newblock Transparent load balancing of mpi programs using ompss-2@ cluster and
  dlb.
\newblock In {\em 51st International Conference on Parallel Processing (ICPP)},
  2022.

\bibitem{zhu2023novel}
Guixun Zhu, Jason Hughes, Siming Zheng, and Deborah Greaves.
\newblock A novel mpi-based parallel smoothed particle hydrodynamics framework
  with dynamic load balancing for free surface flow.
\newblock {\em Computer Physics Communications}, 284:108608, 2023.

\bibitem{mohammed2020two}
Ali Mohammed, Aur{\'e}lien Cavelan, Florina~M Ciorba, Rub{\'e}n~M Cabez{\'o}n,
  and Ioana Banicescu.
\newblock Two-level dynamic load balancing for high performance scientific
  applications.
\newblock In {\em Proceedings of the 2020 SIAM Conference on Parallel
  Processing for Scientific Computing}, pages 69--80. SIAM, 2020.

\bibitem{miller2021dynamic}
Kyle~G Miller, Roman~P Lee, Adam Tableman, Anton Helm, Ricardo~A Fonseca,
  Viktor~K Decyk, and Warren~B Mori.
\newblock Dynamic load balancing with enhanced shared-memory parallelism for
  particle-in-cell codes.
\newblock {\em Computer Physics Communications}, 259:107633, 2021.

\bibitem{liu2012accelerating}
Bin Liu, Dawid Zydek, Henry Selvaraj, and Laxmi Gewali.
\newblock Accelerating high performance computing applications: Using cpus,
  gpus, hybrid cpu/gpu, and fpgas.
\newblock In {\em 2012 13th International Conference on Parallel and
  Distributed Computing, Applications and Technologies}, pages 337--342. IEEE,
  2012.

\bibitem{chen2009high}
Gang Chen, Guobo Li, Songwen Pei, and Baifeng Wu.
\newblock High performance computing via a gpu.
\newblock In {\em 2009 First International Conference on Information Science
  and Engineering}, pages 238--241. IEEE, 2009.

\bibitem{tiwari2015understanding}
Devesh Tiwari, Saurabh Gupta, James Rogers, Don Maxwell, Paolo Rech, Sudharshan
  Vazhkudai, Daniel Oliveira, Dave Londo, Nathan DeBardeleben, Philippe Navaux,
  et~al.
\newblock Understanding gpu errors on large-scale hpc systems and the
  implications for system design and operation.
\newblock In {\em 2015 IEEE 21st International Symposium on High Performance
  Computer Architecture (HPCA)}, pages 331--342. IEEE, 2015.

\bibitem{Callison2022Hybrid}
Adam Callison and Nicholas Chancellor.
\newblock Hybrid quantum-classical algorithms in the noisy intermediate-scale
  quantum era and beyond.
\newblock {\em Phys. Rev. A}, 106:010101, Jul 2022.

\bibitem{de2021materials}
Nathalie~P De~Leon, Kohei~M Itoh, Dohun Kim, Karan~K Mehta, Tracy~E Northup,
  Hanhee Paik, BS~Palmer, Nitin Samarth, Sorawis Sangtawesin, and David~W
  Steuerman.
\newblock Materials challenges and opportunities for quantum computing
  hardware.
\newblock {\em Science}, 372(6539):eabb2823, 2021.

\bibitem{kwon2021gate}
Sangil Kwon, Akiyoshi Tomonaga, Gopika Lakshmi~Bhai, Simon~J Devitt, and
  Jaw-Shen Tsai.
\newblock Gate-based superconducting quantum computing.
\newblock {\em Journal of Applied Physics}, 129(4):041102, 2021.

\bibitem{nielsen2010quantum}
Michael~A Nielsen and Isaac~L Chuang.
\newblock {\em Quantum computation and quantum information}.
\newblock Cambridge university press, 2010.

\bibitem{albash2018adiabatic}
Tameem Albash and Daniel~A Lidar.
\newblock Adiabatic quantum computation.
\newblock {\em Reviews of Modern Physics}, 90(1):015002, 2018.

\bibitem{van2001powerful}
Wim Van~Dam, Michele Mosca, and Umesh Vazirani.
\newblock How powerful is adiabatic quantum computation?
\newblock In {\em Proceedings 42nd IEEE symposium on foundations of computer
  science}, pages 279--287. IEEE, 2001.

\bibitem{aharonov2008adiabatic}
Dorit Aharonov, Wim Van~Dam, Julia Kempe, Zeph Landau, Seth Lloyd, and Oded
  Regev.
\newblock Adiabatic quantum computation is equivalent to standard quantum
  computation.
\newblock {\em SIAM review}, 50(4):755--787, 2008.

\bibitem{farhi2000quantum}
Edward Farhi, Jeffrey Goldstone, Sam Gutmann, and Michael Sipser.
\newblock Quantum computation by adiabatic evolution.
\newblock {\em arXiv preprint quant-ph/0001106}, 2000.

\bibitem{kato1950adiabatic}
Tosio Kato.
\newblock On the adiabatic theorem of quantum mechanics.
\newblock {\em Journal of the Physical Society of Japan}, 5(6):435--439, 1950.

\bibitem{jansen2007bounds}
Sabine Jansen, Mary-Beth Ruskai, and Ruedi Seiler.
\newblock Bounds for the adiabatic approximation with applications to quantum
  computation.
\newblock {\em Journal of Mathematical Physics}, 48(10), 2007.

\bibitem{elgart2012note}
Alexander Elgart and George~A Hagedorn.
\newblock A note on the switching adiabatic theorem.
\newblock {\em Journal of Mathematical Physics}, 53(10), 2012.

\bibitem{stinchcombe1973ising}
RB~Stinchcombe.
\newblock Ising model in a transverse field. i. basic theory.
\newblock {\em Journal of Physics C: Solid State Physics}, 6(15):2459, 1973.

\bibitem{king2022coherent}
Andrew~D King, Sei Suzuki, Jack Raymond, Alex Zucca, Trevor Lanting, Fabio
  Altomare, Andrew~J Berkley, Sara Ejtemaee, Emile Hoskinson, Shuiyuan Huang,
  et~al.
\newblock Coherent quantum annealing in a programmable 2,000 qubit ising chain.
\newblock {\em Nature Physics}, 18(11):1324--1328, 2022.

\bibitem{Crosson2021Diabatic}
E.~J. Crosson and D.~A. Lidar.
\newblock Prospects for quantum enhancement with diabatic quantum annealing.
\newblock {\em Nature Reviews Physics}, 3(7):466--489, Jul 2021.

\bibitem{Callison2021Energetic}
Adam Callison, Max Festenstein, Jie Chen, Laurentiu Nita, Viv Kendon, and
  Nicholas Chancellor.
\newblock Energetic perspective on rapid quenches in quantum annealing.
\newblock {\em PRX Quantum}, 2:010338, Mar 2021.

\bibitem{bertsimas1993simulated}
Dimitris Bertsimas and John Tsitsiklis.
\newblock Simulated annealing.
\newblock {\em Statistical science}, 8(1):10--15, 1993.

\bibitem{yaacoby2022comparison}
Ran Yaacoby, Nathan Schaar, Leon Kellerhals, Oren Raz, Danny Hermelin, and Rami
  Pugatch.
\newblock Comparison between a quantum annealer and a classical approximation
  algorithm for computing the ground state of an ising spin glass.
\newblock {\em Physical Review E}, 105(3):035305, 2022.

\bibitem{starchl2022unraveling}
Elias Starchl and Helmut Ritsch.
\newblock Unraveling the origin of higher success probabilities in quantum
  annealing versus semi-classical annealing.
\newblock {\em Journal of Physics B: Atomic, Molecular and Optical Physics},
  55(2):025501, 2022.

\bibitem{Chancellor2021Range}
Nicholas Chancellor and Viv Kendon.
\newblock Experimental test of search range in quantum annealing.
\newblock {\em Phys. Rev. A}, 104:012604, Jul 2021.

\bibitem{razavy2013quantum}
Mohsen Razavy.
\newblock {\em Quantum theory of tunneling}.
\newblock World Scientific, 2013.

\bibitem{grant2020adiabatic}
Erica~K Grant and Travis~S Humble.
\newblock Adiabatic quantum computing and quantum annealing.
\newblock In {\em Oxford Research Encyclopedia of Physics}. 2020.

\bibitem{scholl2021quantum}
Pascal Scholl, Michael Schuler, Hannah~J Williams, Alexander~A Eberharter,
  Daniel Barredo, Kai-Niklas Schymik, Vincent Lienhard, Louis-Paul Henry,
  Thomas~C Lang, Thierry Lahaye, et~al.
\newblock Quantum simulation of 2d antiferromagnets with hundreds of rydberg
  atoms.
\newblock {\em Nature}, 595(7866):233--238, 2021.

\bibitem{lechner2015quantum}
Wolfgang Lechner, Philipp Hauke, and Peter Zoller.
\newblock A quantum annealing architecture with all-to-all connectivity from
  local interactions.
\newblock {\em Science advances}, 1(9):e1500838, 2015.

\bibitem{ebadi2021quantum}
Sepehr Ebadi, Tout~T Wang, Harry Levine, Alexander Keesling, Giulia Semeghini,
  Ahmed Omran, Dolev Bluvstein, Rhine Samajdar, Hannes Pichler, Wen~Wei Ho,
  et~al.
\newblock Quantum phases of matter on a 256-atom programmable quantum
  simulator.
\newblock {\em Nature}, 595(7866):227--232, 2021.

\bibitem{de2011introduction}
Diego de~Falco and Dario Tamascelli.
\newblock An introduction to quantum annealing.
\newblock {\em RAIRO-Theoretical Informatics and Applications}, 45(1):99--116,
  2011.

\bibitem{rathore2023flame}
Omer Rathore and Salvador Navarro-Martinez.
\newblock Flame dynamics modelling using artificially thickened models.
\newblock {\em Flow, Turbulence and Combustion}, pages 1--31, 2023.

\bibitem{bell2012boxlib}
J~Bell, A~Almgren, V~Beckner, M~Day, M~Lijewski, A~Nonaka, and W~Zhang.
\newblock Boxlib user’s guide.
\newblock {\em github. com/BoxLib-Codes/BoxLib}, 2012.

\bibitem{zhang2016boxlib}
Weiqun Zhang, Ann Almgren, Marcus Day, Tan Nguyen, John Shalf, and Didem Unat.
\newblock Boxlib with tiling: An adaptive mesh refinement software framework.
\newblock {\em SIAM Journal on Scientific Computing}, 38(5):S156--S172, 2016.

\bibitem{almgren2010castro}
Ann~S Almgren, Vincent~E Beckner, John~B Bell, MS~Day, Louis~H Howell,
  CC~Joggerst, MJ~Lijewski, Andy Nonaka, M~Singer, and Michael Zingale.
\newblock Castro: A new compressible astrophysical solver. i. hydrodynamics and
  self-gravity.
\newblock {\em The Astrophysical Journal}, 715(2):1221, 2010.

\bibitem{nonaka2010maestro}
Andrew Nonaka, AS~Almgren, JB~Bell, MJ~Lijewski, CM~Malone, and M~Zingale.
\newblock Maestro: An adaptive low mach number hydrodynamics algorithm for
  stellar flows.
\newblock {\em The Astrophysical Journal Supplement Series}, 188(2):358, 2010.

\bibitem{pau2009parallel}
George~SH Pau, Ann~S Almgren, John~B Bell, and Michael~J Lijewski.
\newblock A parallel second-order adaptive mesh algorithm for incompressible
  flow in porous media.
\newblock {\em Philosophical Transactions of the Royal Society A: Mathematical,
  Physical and Engineering Sciences}, 367(1907):4633--4654, 2009.

\bibitem{day2000numerical}
Mark~S Day and John~B Bell.
\newblock Numerical simulation of laminar reacting flows with complex
  chemistry.
\newblock {\em Combustion Theory and Modelling}, 4(4):535, 2000.

\bibitem{karp2010reducibility}
Richard~M Karp.
\newblock {\em Reducibility among combinatorial problems}.
\newblock Springer, 2010.

\bibitem{schaller2023swift}
Matthieu Schaller, Josh Borrow, Peter~W Draper, Mladen Ivkovic, Stuart
  McAlpine, Bert Vandenbroucke, Yannick Bah{\'e}, Evgenii Chaikin, Aidan~BG
  Chalk, Tsang~Keung Chan, et~al.
\newblock Swift: A modern highly-parallel gravity and smoothed particle
  hydrodynamics solver for astrophysical and cosmological applications.
\newblock {\em arXiv preprint arXiv:2305.13380}, 2023.

\bibitem{schaller2016swift}
Matthieu Schaller, Pedro Gonnet, Aidan~BG Chalk, and Peter~W Draper.
\newblock Swift: Using task-based parallelism, fully asynchronous
  communication, and graph partition-based domain decomposition for strong
  scaling on more than 100,000 cores.
\newblock In {\em Proceedings of the platform for advanced scientific computing
  conference}, pages 1--10, 2016.

\bibitem{karypis1997metis}
George Karypis and Vipin Kumar.
\newblock Metis: A software package for partitioning unstructured graphs,
  partitioning meshes, and computing fill-reducing orderings of sparse
  matrices.
\newblock 1997.

\bibitem{cai2014practical}
Jun Cai, William~G Macready, and Aidan Roy.
\newblock A practical heuristic for finding graph minors.
\newblock {\em arXiv preprint arXiv:1406.2741}, 2014.

\bibitem{zbinden2020embedding}
Stefanie Zbinden, Andreas B{\"a}rtschi, Hristo Djidjev, and Stephan Eidenbenz.
\newblock Embedding algorithms for quantum annealers with chimera and pegasus
  connection topologies.
\newblock In {\em International Conference on High Performance Computing},
  pages 187--206. Springer, 2020.

\bibitem{barbosa2021optimizing}
Aaron Barbosa, Elijah Pelofske, Georg Hahn, and Hristo~N Djidjev.
\newblock Optimizing embedding-related quantum annealing parameters for
  reducing hardware bias.
\newblock In {\em Parallel Architectures, Algorithms and Programming: 11th
  International Symposium, PAAP 2020, Shenzhen, China, December 28--30, 2020,
  Proceedings 11}, pages 162--173. Springer, 2021.

\bibitem{barbosa2021using}
Aaron Barbosa, Elijah Pelofske, Georg Hahn, and Hristo~N Djidjev.
\newblock Using machine learning for quantum annealing accuracy prediction.
\newblock {\em Algorithms}, 14(6):187, 2021.

\bibitem{cavallaro2022hybrid}
Gabriele Cavallaro, Morris Riedel, Thomas Lippert, and Kristel Michielsen.
\newblock Hybrid quantum-classical workflows in modular supercomputing
  architectures with the {JULICH} unified infrastructure for quantum computing.
\newblock In {\em IGARSS 2022-2022 IEEE International Geoscience and Remote
  Sensing Symposium}, pages 4149--4152. IEEE, 2022.

\bibitem{lippert2022perspectives}
T~Lippert and K~Michielsen.
\newblock Perspectives of quantum computing at the {JULICH} supercomputing
  centre.
\newblock 2022.

\end{thebibliography}
\end{document}